\documentclass[12pt,fleqn,a4paper]{article}
\pdfoutput=1
\usepackage{graphicx}
\usepackage{epsfig,cite}
\usepackage{amssymb}
\usepackage{amsmath}
\usepackage{color}
\usepackage{amstext,alltt,setspace}
\usepackage{amsbsy}
\usepackage{array}
\usepackage{booktabs, makecell}
\usepackage{hyperref}
\usepackage{slashed}
\usepackage{url}
\usepackage{geometry}
\providecommand{\pDa}[1]{{
\left(
\frac{1}{1-#1}
\right)_+
}}

\newcommand{\msbar}{$\overline{\rm { MS}}$}
\newcommand{\mmsbar}{{\overline{\rm {MS}}}}
\newcommand{\NS}{{\rm NS}}

\def\toinf#1{\mathrel{\mathop{\sim}\limits_{\scriptscriptstyle
{#1\rightarrow\infty }}}}
\begin{document}
\begin{flushleft}
  \includegraphics[width=.3\linewidth,clip]{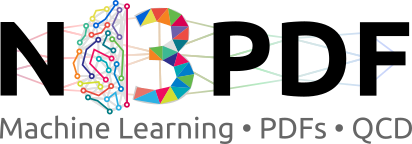}
\end{flushleft}
\vspace{-2.5cm}
\begin{flushright}
  \includegraphics[width=.15\textwidth]{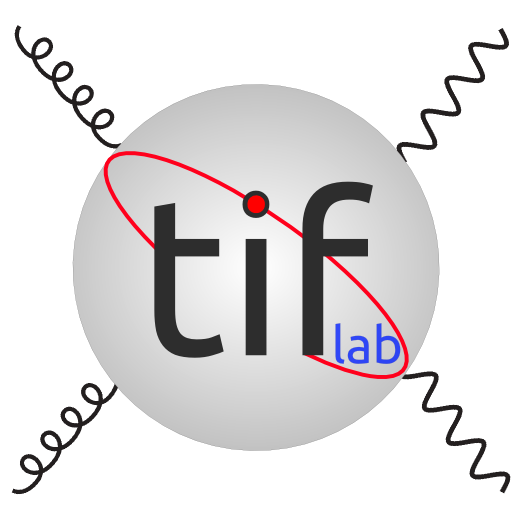}\\
  TIF-UNIMI-2020-9\\
\end{flushright}

\vspace*{.2cm}

\begin{center}
  {\Large \bf{Can \msbar{} parton distributions be negative? }}
\end{center}

\vspace*{.7cm}

\begin{center}
  Alessandro Candido, Stefano Forte and Felix Hekhorn
  \vspace*{.2cm}

{  \noindent
      {\it
        Tif Lab, Dipartimento di Fisica, Universit\`a di Milano and\\ 
        INFN, Sezione di Milano,
        Via Celoria 16, I-20133 Milano, Italy}}

      \vspace*{3cm}

      {\bf Abstract}
\end{center}

{\noindent It is common lore that Parton Distribution Functions (PDFs) in the \msbar{}
  factorization scheme  can become negative beyond leading order due
  to the collinear subtraction which is needed in order to define
  partonic cross sections. We
  show that this is in fact not the case and next-to-leading order (NLO)
  \msbar{} PDFs are actually
  positive in the perturbative regime. In order to prove this, we
  modify the subtraction prescription, and perform the collinear
  subtraction in such a way that partonic cross sections remain
  positive. This defines  a factorization scheme in which PDFs
  are positive.
  We then show
 that positivity of the PDFs is preserved when transforming from this
scheme to \msbar{}, provided only the strong coupling is in the 
perturbative regime, such that the NLO scheme change is smaller than
the LO term.}

\pagebreak

\section{Parton distributions from the parton model to QCD}
\label{sec:intro}
In the naive parton model parton distribution functions (PDFs) are
probability densities for a parton to carry a given fraction of their
parent hadron's momentum. This simple picture remains true at leading
order (LO) in QCD where each PDF is proportional to a
physically measurable (in principle) cross section: it can
consequently still be viewed as a probability and it is
non-negative. We will henceforth refer
to the property of being non-negative as
``positivity'' (while really meaning ``positive-semidefiniteness'').

As is well known~\cite{Altarelli:1998gn}, positivity of the PDFs
is in general violated at higher
perturbative orders, where physical cross sections are the convolution
of the PDFs with a partonic cross section. Partonic cross sections are
beset by collinear singularities, whose subtraction and
factorization into the PDF spoils in general the 
separate positivity of the subtracted partonic cross sections and of
the
 PDFs. Therefore,
the positivity condition on PDFs beyond LO becomes a
positivity constraint of physical
observables~\cite{Altarelli:1998gn,Forte:1998kd}. When determining
PDFs from data, this constraint must be imposed on physical
observables, rather than PDFs themselves: for instance by using
Lagrange multipliers, or pseudodata for suitable
positivity observables, such as for example hypothetical deep-inelastic
structure functions corresponding to gauge bosons that couple to only
one quark flavor~\cite{Ball:2008by}.

These positivity constraint 
may have a significant impact on
PDF determination, especially in regions where there are little or no direct
constraints coming from experimental data. For example, in a recent
study of the strange PDF~\cite{Faura:2020oom} it was found that
adding to the dataset a positivity constraint for the charm structure
function $F_2^c$ significantly impacts the flavor decomposition of the
quark sea in the
large $x$ region. However, adding positivity constraints in this way,
as constraints on pseudo-observables, is cumbersome from a methodological point
of view, and theoretically not satisfactory. Indeed, as  the target
accuracy of the PDFs increases, and PDFs are used more and more
for new physics studies and searches in regions in which the PDFs are
experimentally 
unconstrained, it becomes necessary to enforce 
an increasingly
elaborate set of positivity
constraints~\cite{Ball:2010de,Ball:2014uwa} for
a set of suitably chosen and tuned pseudo-observables. This poses
obvious problems of fine-tuning. In fact,
universality of PDFs suggests that  positivity constraints
should be imposed in a process-independent way, without having to rely
on a specific choice  of processes, and therefore, that it ought to be
possible to impose the constraint at the level of PDFs.

Here, we address this issue head-on by constructing a subtraction scheme in
which PDFs are positive, and which we
refer to as a ``positive'' factorization scheme. We do this by
studying the way collinear subtraction is performed in the \msbar{}
scheme, and showing that negative partonic
cross sections arise as a consequence of
over-subtraction of a positive contribution\footnote{A subtlety is related to the fact that
  generally partonic cross sections are the sum of an ordinary
  function of the scaling variable, and a distribution localized at
  the kinematic threshold of the scaling variable. Here, by ``negative
  cross section'' we mean that the function (i.e., non-distributional
  part of the cross section) is negative. For positivity to hold, the
  distributional part must also be positive in the sense that it gives
a positive result when integrated over a positive test function. As we
shall see below, this condition turns out to be automatically satisfied
in \msbar{} and related schemes.}.
This is chiefly due to the fact that the subtraction is
performed at a scale which, as the  kinematic threshold for production
of the final state is approached,
is higher than the actual physical scale;
and also, in gluonic
channels, due to the way
the $d$-dimensional averaging over gluon polarizations is treated in
dimensional regularization.

Once these effects are taken into account
it is possible to formulate a subtraction prescription such that
partonic cross sections remain
positive. Effectively, this choice of subtraction corresponds to a ``physical''
scheme, in which the scale choice is directly related to the scale of
parton radiation.
Of course, the
positive factorization scheme is not unique, since any
further scheme change through a matrix with positive entries
(``positive matrix'', henceforth) would leave 
the partonic cross sections positive.  It is then possible to show that there exist
schemes in which PDFs also remain positive, so that the positive
hadronic cross section is obtained by convoluting positive
partonic cross sections with positive PDFs.

The availability of positive schemes can be
advantageous in the 
context of PDF determination. Indeed, if PDFs are
parametrized in the positive scheme, positivity can be enforced by
choice of parametrization. Results in the commonly used \msbar{} scheme
can then be obtained by scheme transformation.

However, perhaps surprisingly, this turns out not to be necessary:
indeed, using the explicit form of the scheme change
matrix from the positive scheme to \msbar{} it is easy to prove that in
the perturbative region  PDFs remain positive in
\msbar{}. Hence, the common lore that \msbar{} PDFs might be negative
beyond LO
turns out to be incorrect. Positivity of the fitted PDFs can then be imposed
using the standard methodology in the
\msbar{} scheme.

The paper is organized as follows. In Section~\ref{sec:subtr} we show
how negative partonic cross sections arise due to over-subtraction. We
start with the   prototypical case of deep-inelastic scattering (DIS): we
review the computation of coefficient functions at next-to-leading
order (NLO); we show how over-subtraction arises in the \msbar{} scheme
and how it can be fixed by choosing a suitable subtraction
prescription; we then discuss how this works in the general case of
hadronic processes, where we can define a ``positive'' subtraction prescription
which preserves
positivity of all partonic cross sections.
In Sect.~\ref{sec:scheme} we then turn to
positivity of PDFs: first, we use our positive subtraction
prescription to define a positive factorization scheme; then we show
how positivity of PDFs is preserved in this factorization scheme; and
finally by studying the transformation from the positive scheme to  \msbar{}
we prove that positivity is preserved in \msbar{} in the perturbative
regime. The bulk of our discussion will be at NLO, and its validity
beyond NLO will be addressed in the end of Sect.~\ref{sec:scheme}.

\section{Positivity of partonic cross sections}
\label{sec:subtr}

QCD factorization allows expressing
physical cross sections $\sigma$  as convolutions of
partonic cross sections with parton distributions
$f_i$. In the prototypical case of DIS
the cross section is expressed
in terms of hadronic structure functions $F(x,Q^2)$, which are then
factorized in terms of  parton-level structure
functions, called coefficient functions $C_i$:
\begin{equation}
  \label{eq:masterfac}
 \frac{1}{x} F(x,Q^2)=\sum_{i} e^2_i C_i \otimes f_i ,
\end{equation}
where the sum runs over all parton species, $e_i$ are quark
electric charges, or the sum over all electric charges for the gluon,
(for photon-induced DIS, and more in general electroweak charges),
$\otimes$ denotes convolution, and we refer to
  Ref.~\cite{Ellis:1991qj} for
  notations and conventions. The convolution in
  Eq.~(\ref{eq:masterfac}) links the three a priori physically
  distinct scaling variables on which respectively the physical
  observable $F$, the partonic cross-section $C$ and the PDF $f$
  depend. In the sequel, for clarity, we will denote with $x$ the
  physically observable variable  (Bjorken-$x$ for DIS, or the scaling
  variable in hadronic collisions), with $z$ the variable on which the
  coefficient function depends, and with $\xi$ the PDF momentum
  fraction. Of course, Mellin transformation turns the convolution
  into an ordinary product and upon transformation all these variables
  are mapped onto the same $N$ variable.

  At LO all factors on the
  right-hand side of
  Eq.~(\ref{eq:masterfac}) are manifestly positive. Indeed, the partonic
  cross sections (which for DIS at LO are trivial) are defined as the square modulus of amplitudes. The
  PDFs in turn are defined as operator matrix elements
  which can be interpreted as probability
  distributions~\cite{Collins:1981uw,Curci:1980uw}: for quark
  PDFs~\cite{Collins:1981uw}  
  \begin{equation}
    \label{eq:quark}
f_i(\xi)=\frac{1}{4\pi}\int dy^- e^{-i \xi P^+ y^-}\langle P| 
\bar \psi_i(0,y^-,\vec 0_T)\gamma^+ {\cal P} \exp\left[ i g_s \int_0^{y^-} d
  \bar y^- A_a^+(0,\bar y^-,\vec 0_T)\frac{1}{2} \lambda_a\right] \psi_i(0) |P\rangle ,
\end{equation}
where ${\cal P}$ denotes path-ordering; $P$ is the four-momentum of
the parent hadron in light-cone components and $g_s$ is the strong
coupling, with analogous expressions for antiquarks and
gluons~\cite{Collins:1981uw}.  It can be shown (see
e.g.\ Sect.~6.7 of Ref.~\cite{Collins:2011zzd}) that the expression
Eq.~(\ref{eq:quark}) is a number density, and as such before
subtraction of divergences it is positive.

Beyond LO, besides ultraviolet renormalization, both the
PDFs and the partonic cross section are beset by collinear
singularities which can be factored into the PDF. Before
factorization the PDF is a ``bare'' probability density $f_i^{(0)}$~\cite{Collins:2011zzd}, while after
factorization it is a
renormalized PDF  $f_i$
\begin{equation}
  \label{eq:pdffac}
f_i =\sum_{j} Z^S_{ij}\otimes f_j^{(0)}. 
\end{equation}
In operator language, the  factor $Z^S_{ij}$ is a multiplicative
renormalization of the operator Eq.~(\ref{eq:quark}), which admits a
perturbative expansion
\begin{equation}\label{eq:pertz}
Z^S_{ij}(Q^2)=\delta_{ij}+\frac{\alpha_s}{2\pi}
\delta^S_{ij}(Q^2)+O(\alpha_s^2),
\end{equation}
where $\delta^S_{ij}$ is a counterterm which diverges after
regularization is removed, the superscript $S$ denotes the fact
that the finite part of the counterterm depends on the choice of a
particular subtraction scheme $S$, and regularization induces a
dependence of the counterterm and thus of the renormalization constant
on scale.

The counterterm can be determined
in a standard way by taking the matrix element of the
operator in a state in which the right-hand side of
Eq.~(\ref{eq:quark}) is perturbatively computable, such as a free
state of a parton $i$, in which the PDF for finding a parton $j$ is
trivially
\begin{equation} \label{eq:freef}
  f_j^{i\,(0)}(\xi)=\delta_{ij}\delta(1-\xi),
\end{equation}
imposing a
renormalization condition and finally removing the regulator. In practice, this
is most easily done~\cite{Collins:2011zzd,Curci:1980uw} by introducing
a probe that couples to the free 
quark, so for instance computing the structure function
Eq.~(\ref{eq:masterfac}) for deep-inelastic scattering off a free
quark. This is the strategy that we will follow in this section, where
such a computation will be performed explicitly in a way that fully
determines the factorization scheme, both in the \msbar{} and in our
new positive schemes.

The factorization argument then works as follows. The $d$-dimensional structure
function Eq.~(\ref{eq:masterfac}) is
written as
\begin{align} \label{eq:refaca}
 \frac{1}{x} F_i(x,Q^2,\epsilon)&=\sum_{j} e^2_j C_j \otimes f_j^{i\,(0)} \\
&=\sum_{j} e^2_j C_j^S \otimes f_j^{i\,S};\qquad\quad d=4-2\epsilon,
\label{eq:refacb}
\end{align}
and computed by taking in turn  the incoming parton to be each of the
parton species, i.e. using Eq.~(\ref{eq:freef}).
  Of course, the structure
function on the l.h.s.\ then reduces to the
unsubtracted, regularized coefficient
function, which is essentially the cross-section for scattering off
the given incoming free parton. The counterterm is defined by imposing the
cancellation of the singularity. Up to NLO, assuming a free incoming
parton according to Eq.~(\ref{eq:freef}), substituting in
Eqs.~(\ref{eq:refaca}-\ref{eq:refacb}) the
perturbative expression Eq.~(\ref{eq:pertz}) of the 
renormalization factor Eq.~(\ref{eq:pdffac}), and assuming a
perturbative expansion of the coefficient functions of the form
\begin{equation}\label{eq:perc}
  C_i(z,Q^2)= C^{(0)}_i(z,Q^2)+\frac{\alpha_s}{2\pi} C^{(1)}_i(z,Q^2)+O(\alpha_s^2)
\end{equation} 
one gets
\begin{equation}\label{eq:renorm}
    C_i^S(z,Q^2,\epsilon) = 
    {C^{(1)}_{i}}(z,Q^2,\epsilon)
    -\delta_{qi}^S(z,Q^2,\epsilon)\,,
\end{equation}
where $q$ denotes a quark parton.
Note that, up to NLO, imposing finiteness of the DIS structure
functions fixes the renormalization in the quark sector because DIS is
a probe that only couples to quarks at leading order.

The advantage of determining the counterterms in this way, as opposed
to performing a direct computation of the current matrix element
Eq.~(\ref{eq:quark}) is that in  operator matrix
elements all divergences appear as ultraviolet, while, when computing a
structure function for an incoming free parton  (or, more
generally, a generic partonic cross-section), collinear singularities
come from the infrared region of integration over
transverse momenta. Hence, one may compute the relevant
cross-section using renormalized perturbation theory (i.e., with
counterterms already included in the Lagrangian). The only divergences
are then of collinear and infrared origin.
The regularized partonic cross-section is
then finite if the computation is
performed with $\epsilon<0$, and it enjoys the positivity
properties of a standard 
cross-section. This property will be crucial in the argument presented
 below.

After the subtraction Eq.~(\ref{eq:renorm}), the partonic cross-section
(coefficient function) is
finite in the $\epsilon\to0$ limit, so one may define the
four-dimensional coefficient function as
\begin{equation}\label{eq:crenorm}
    {C^{(1)}_{i}}^S(z) = \lim_{\epsilon\to0^-}\left( {C^{(1)}_{i}}(z,Q^2,\epsilon)
    -\delta_{qi}^S(z,Q^2,\epsilon)\right),
\end{equation}
where $\epsilon\to 0^-$ denotes the fact that the limit is taken from
below, as discussed above. Note that the four-dimensional coefficient
function function can depend only on $z$ for dimensional reasons,
while the $d$ dimensional one  also depends on $Q^2$ through the
combination $\frac{Q^2}{\mu^2}$, where $\mu^2$ is the scale of
dimensional regularization. That this subtraction is always possible
is the content of factorization
theorems~\cite{Collins:2011zzd,Curci:1980uw}. The universal
(i.e.\ process-independent) nature of the collinear singularities
ensure that the renormalization conditions on parton distributions,
defined as operator matrix elements Eq.~(\ref{eq:quark})
without reference to any specific
process, may be determined by the computation of a particular process
or set of processes as discussed here.

The finite part of the
subtraction is arbitrary and it defines the factorization scheme $S$. In
\msbar{} it turns out that in some partonic subchannels the
subtracted cross section can be negative:
effectively, negative finite parts are factored away from the
regularized cross sections, and into the PDFs. These can then
also become negative, though whether this happens or not depends on
the relative weight of the various subchannels.
On the other hand, the residue of
the collinear pole is universal---it is given by
process-independent splitting functions---and this makes it possible
to define its subtraction in a way that preserves positivity of the
partonic cross section at the regularized level. If all contributions
which are factored away from the partonic cross section and into the
PDF remain positive, then the latter also stays positive.

Having explained the general strategy, we now implement it
explicitly. 
We first discuss DIS structure functions.
We  then turn to hadronic processes, both quark-induced and gluon induced.

\subsection{Deep-inelastic coefficient functions}
\label{sec:discf}
At NLO, photon-induced DIS proceeds through the two
sub-processes $q+\gamma^*\to X$ and $g+\gamma^*\to X$, in such a way
that the contribution of each quark or antiquark flavor to the
structure function $F_2$ can be written as:
\begin{equation}\label{eq:f2}
 \frac{1}{x} F_2(x,Q^2)= e_q^2 \left[q
 +\frac{\alpha_s}{2\pi}\left( C^{(1)}_q \otimes q+ C^{(1)}_g \otimes
 g\right)\right](Q^2)\,,
 \end{equation}
where $e_q$ is the electric charge of the quark, on the right-hand
side we have omitted the $x$ dependence which arises from the
convolution,
and the
generalization to $Z$- and $W$-induced DIS is trivial.

\begin{figure}[t]
  \begin{center}
    \includegraphics[width=0.8\linewidth]{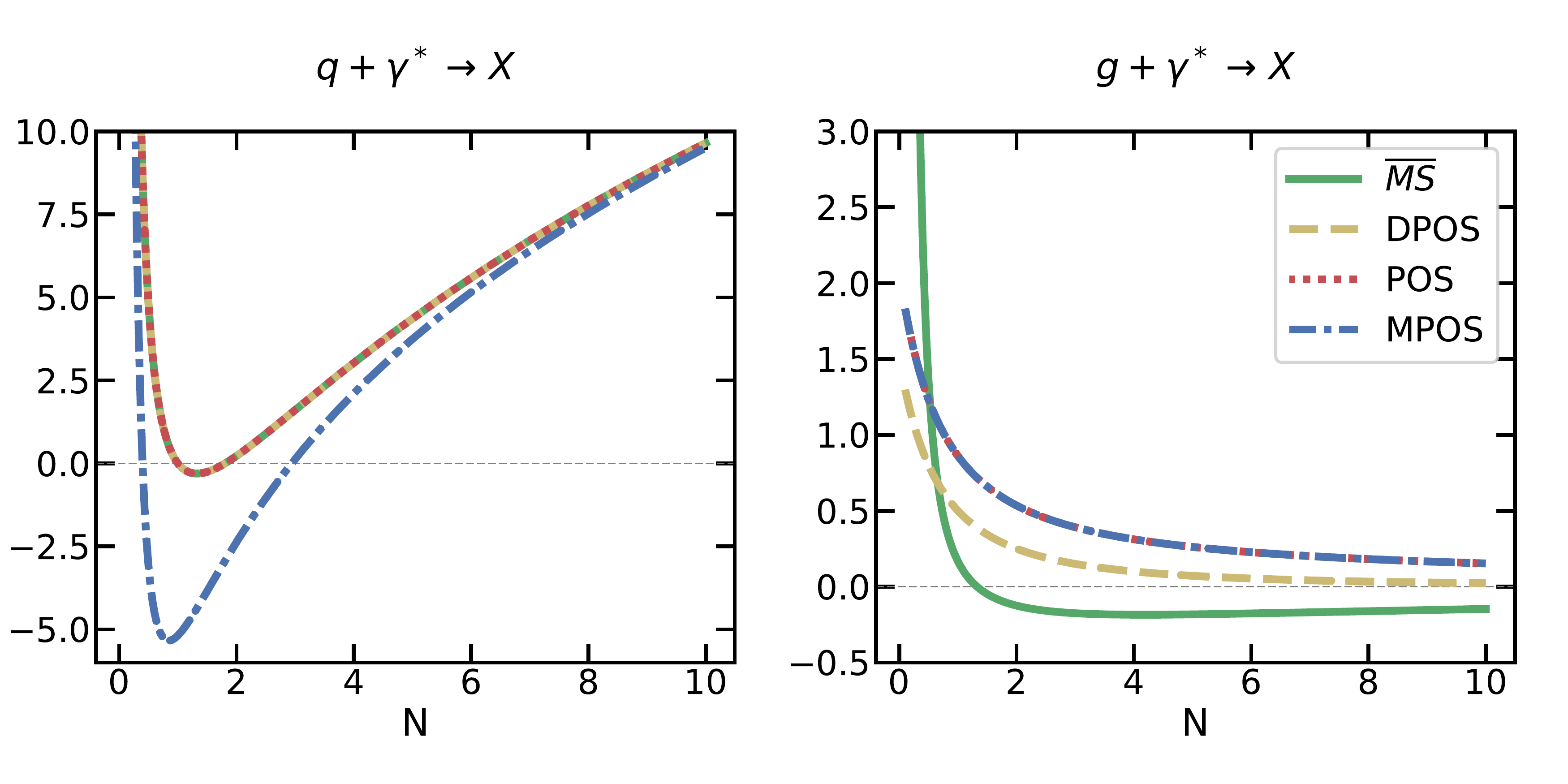}
    \caption{\small Mellin-space NLO contributions to deep-inelastic coefficient
      functions. The quark (left) and gluon (right) coefficient
      functions, respectively $C_q^{(1)}$ and $C_g^{(1)}$, Eq.~\ref{eq:f2}, are
      shown. The DPOS
      scheme is defined in Eqs.~(\ref{eq:count},\ref{eq:countq}), the POS
      scheme is defined in Eqs.~(\ref{eq:counthqq}-\ref{eq:counthqg}),
      and the MPOS scheme
      in Eqs.~(\ref{eq:mposqq}-\ref{eq:mposqg}).
      Results are shown in the \msbar{} and DPOS, POS and MPOS
      schemes. For  $C_q^{(1)}$ \msbar{}, DPOS and POS
      coincide, and the two curves shown correspond, from top to
      bottom, to \msbar{} and MPOS; for $C_g^{(1)}$ POS and MPOS
      coincide and the three
      curves correspond,
      from  bottom to top, to \msbar{}, DPOS and POS.
    \label{fig:dis} }
  \end{center}
\end{figure}
  The
  \msbar{} NLO contributions to the coefficient functions
  $C_q$ and $C_g$ are shown in Fig.~\ref{fig:dis} in Mellin space,
  where the convolution becomes an ordinary product. The
  Mellin space plot is especially transparent since the $x$-space
  cross section is found to high accuracy by computing the inverse
  Mellin transform in the saddle-point
  approximation~\cite{Bonvini:2012an}: hence, the physical  $x$-space
  cross section is just the product of the Mellin-space coefficient
  function and PDF evaluated at the value of $N$ corresponding to the
  saddle for given kinematics. 
  It is clear  from Fig.~\ref{fig:dis} that at
  large $N$ the gluon coefficient function is negative on the real
  axis: hence, the $x$-space coefficient function must also be
  negative because its real moments are negative. This shows
 that a negative contribution has been factored from the coefficient
 function into the PDF.

\subsubsection{Over-subtraction and the off-diagonal coefficient function}
\label{sec:offdiag}

In order to understand what is going on, we look at the dimensionally
regularized, unsubtracted gluon coefficient function:
\begin{equation}\label{eq:cgeps}
C^{(1)}_{g}(z,Q^2,\epsilon) = \frac{ \Gamma(-\epsilon)
  \left(\frac{\mu_D^2}{\pi\mu^2}\right)^{-\epsilon} \left[8P_{qg}(z)-16 T_R \epsilon (3
    -\epsilon(2 -\epsilon) )  \right]  }{16\pi (2 - 2\epsilon) \Gamma (3 - 2 \epsilon)}\,,
\end{equation}
where
\begin{equation}\label{eq:mud}
  \mu_D^2=\frac{s}{4}=\frac{Q^2(1-z)}{4z},
\end{equation} 
and  $s=\frac{Q^2(1-z)}{z}$ is the center-of-mass energy of the
$\gamma^* q$ collision. 
Note that in order to regulate the collinear singularity it is
necessary to choose $\epsilon<0$; it then follows that  as $\epsilon$
goes to zero from below,
$\Gamma(-\epsilon)>0$ and
the unsubtracted coefficient function, Eq.~(\ref{eq:cgeps}), is positive as it ought to
be.

The subtracted \msbar{} coefficient function is then given by
\begin{align}\label{eq:cgren}
{C^{(1)}_{g}}^{\mmsbar}(z) &= \lim_{\epsilon\to
  0^-}\left[C_{g}^{(1)}(z,Q^2,\epsilon) - \left(\frac{Q^2}{4\pi
      \mu^2}\right)^{-\epsilon} \left(-\frac 1
    {\epsilon}+\gamma_E\right) P_{qg}(z)\right]\\
    &= P_{qg}(z) \left( \ln\left(\frac{1-z}{z}\right) - 4 \right) + 3
    T_R \, ,\label{eq:cgnlo}
\end{align}
where  $\epsilon\to 0^-$ denotes the fact that the limit should be
taken from below, because the collinear singularity is regulated with
$\epsilon<0$. 
The $P_{qg}$ splitting function is positive for all $z$, so for
$z>\frac{1}{2}$ the log  becomes negative and at large $z$ the
coefficient function is negative.

Comparing  Eqs.~(\ref{eq:cgeps}-\ref{eq:cgren}) immediately reveals
what happened:  the regularized coefficient function contains a term
\begin{equation}\label{eq:logexp}
  \left(\frac{s/4}{\pi\mu^2}\right)^{-\epsilon}=1-\epsilon \ln \left(\frac{Q^2(1-z)/z}{4\pi\mu^2}\right),
\end{equation}
but in the collinear subtraction $\ln \frac{Q^2}{4\pi\mu^2}$ has been
subtracted instead. For $z>\frac{1}{2}$, $s<Q^2$ this
amounts to over-subtracting, at the larger scale $Q^2$ instead of the
smaller physical scale $s$. The physical origin of this contribution,
and the reason for the mismatch are easy to trace.

Namely, this is the
contribution coming from quark emission from the incoming gluon line,
and the singularity is due to the collinear singular integration over
the transverse momentum of the emitted quark, as revealed by the fact
that it is proportional to the corresponding $P_{qg}$ splitting
function. The argument of the ensuing collinear log is set by the
upper limit of the transverse momentum integration $k_T^{\rm max}$,
which for a $2\to2$ process with massless particles in the final state
is $k_T^{\rm max}=\frac{s}{4}$. In \msbar{}
the collinear subtraction is performed at the scale $Q^2$,
hence leading to the over-subtraction that we observed, and producing
a contribution to the coefficient function which is logarithmically
enhanced in the threshold $z\to1$ limit.

Therefore, this contribution has the same origin as the soft (Sudakov)
logarithms 
which are resummed to all orders
when performing threshold
resummation~\cite{Catani:1989ne,Sterman:1986aj}, except that 
in soft resummation the splitting function is evaluated in the $z\to1$
limit, and the factor of $\frac{1}{z}$ in the argument of the log is
neglected.
In fact, threshold resummation can be
obtained by identifying (and then
renormalization-group improving)
\begin{equation}\label{eq:ktmax}
 |k_T^{\rm max,\,DIS}|^2=\mu_D^2
\end{equation}
(with $\mu_D^2$ given by Eq.~(\ref{eq:mud}))
  as the physical
scale in the soft limit~\cite{Forte:2002ni}. The over-subtraction is then
simply the manifestation of the well-known fact that, in the \msbar{} scheme,
threshold logs beyond the first are factored in the coefficient
function, and not in the PDF~\cite{Albino:2000cp}. Indeed, alternative
factorization schemes in which these logs are instead included in the
PDF have been proposed, in particular  the Monte Carlo
factorization scheme of Ref.~\cite{Jadach:2016acv}. Note, however, that
radiation in off-diagonal parton channels is power-suppressed in the
threshold limit, and indeed this contribution is proportional to
$\ln(1-z)$, which in Mellin space behaves as $\frac{\ln N}{N}$. This is to be
contrasted with the $\left(\frac{\ln(1-z)}{1-z}\right)_+$ behavior, corresponding to 
$\ln^2 N$, found in diagonal channels, as we shall discuss in
Sect.~\ref{sec:diag} below. Hence, while it has the same origin, this
contribution is not among those included in standard leading-power
threshold resummation. 

In conclusion,  in order to
restore positivity it is  sufficient to perform the collinear
subtraction at the scale $\mu_D^2=s/4$, Eq.~(\ref{eq:ktmax}).
There is a further subtlety, however. Namely, the factor $2-2\epsilon$
in the denominator of Eq.~(\ref{eq:cgeps}) is the average over the
polarization states of the incoming gluon. Therefore, it should be
viewed as an overall prefactor which is common to both the
unsubtracted and subtracted coefficient function, and thus  must be
included in the subtraction term. Because it interferes with a
$-\frac{1}{\epsilon}$ pole, not including it, as in \msbar{}, leads to
over-subtraction: 
the collinear singularity is regulated with
$\epsilon<0$, so $\frac{1}{1-\epsilon}<1$.  

Therefore, we define a modified positivity subtraction as
\begin{align}\label{eq:cgrenpos}
  {C^{(1)}_{g}}^{\rm DPOS}(z) &= \lim_{\epsilon\to0^-}
  \left[{C^{(1)}_{g}}(z,Q^2,\epsilon) - \frac{1}{1-\epsilon}\left(\frac{\mu_D^2}{\pi\mu^2}\right)^{-\epsilon} \left(-\frac 1
    {\epsilon}+\gamma_E\right) P_{qg}(z)\right]\\\label{eq:cgrenpos1}
    &= 3\left[T_R- P_{qg}(z)\right].
\end{align}
Note that the normalization of the prefactor is fixed by the
requirement of cancellation of the pole.
The coefficient function of Eq.~(\ref{eq:cgrenpos}) is positive definite,
as it is easy to check explicitly. Its Mellin-space form is also shown
in Fig.~\ref{fig:dis}, and it is manifestly positive.

We
can rewrite the subtraction which relates  the regularized
coefficient function, Eq.~(\ref{eq:cgeps}), to its renormalized
counterparts Eqs.~(\ref{eq:cgren},\ref{eq:cgrenpos}) in terms of
counterterms according to Eq.~(\ref{eq:crenorm}), where now
$S=\mmsbar$, DPOS.
We then have 
\begin{align}\label{eq:count}
  &   {C^{(1)}_{g}}^{\rm DPOS}(z) =  {C^{(1)}_{g}}^{\mmsbar}(z) -
  K_{qg}^{\rm DPOS}(z) \,,\\ \label{eq:count1}
  &  K_{qg}^{\rm DPOS}(z)=\delta_{qg}^{\mmsbar}-\delta_{qg}^{\rm DPOS}=  P_{qg}(z)\left[\ln\left(\frac{1-z}{z}\right) - 1\right].
\end{align}

\begin{figure}[t]
  \begin{center}
    \includegraphics[width=0.8\linewidth]{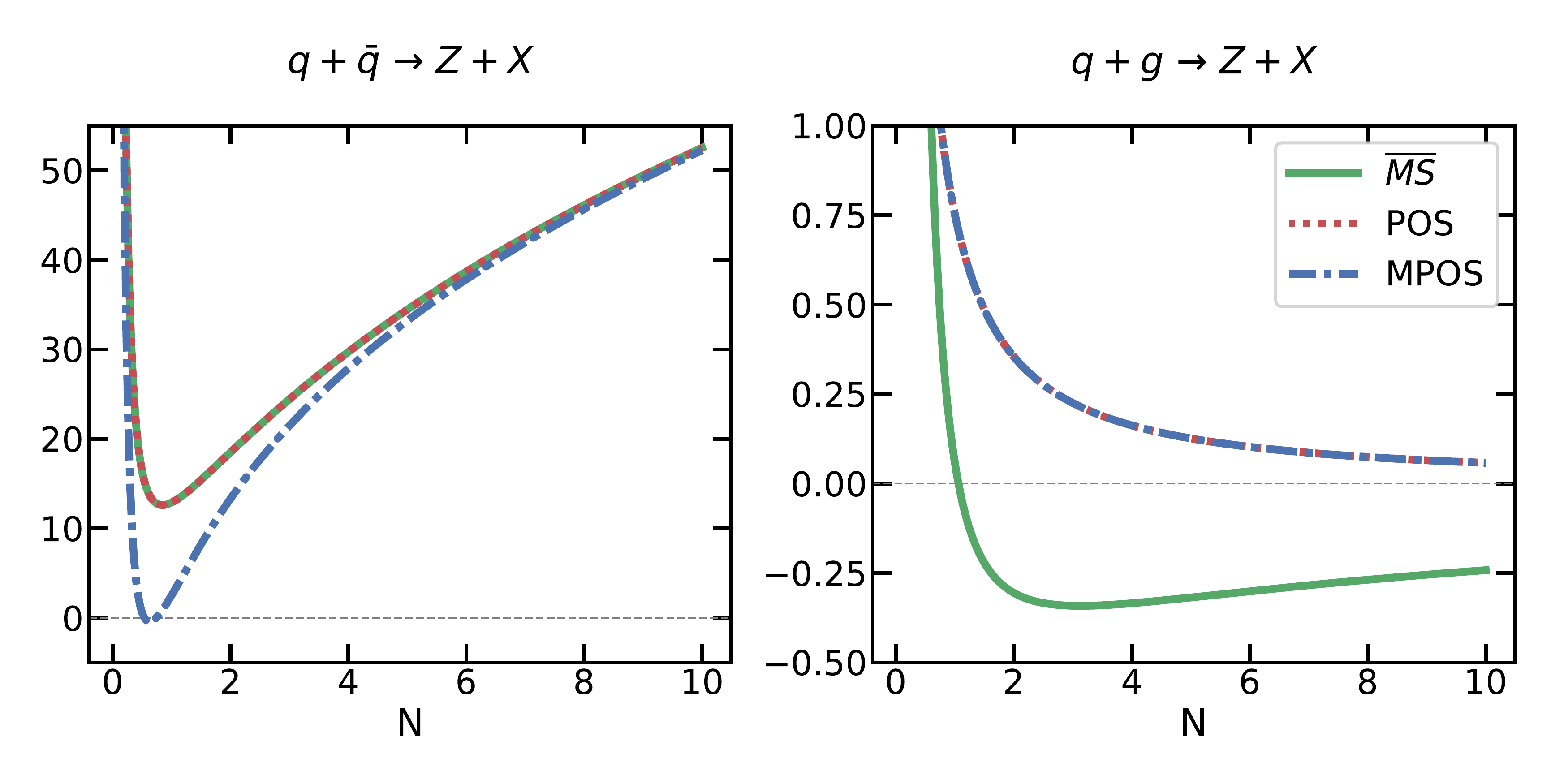}
    \caption{\small Mellin-space NLO contributions to Drell--Yan coefficient
      functions. The quark (left) and gluon (right) coefficient
      functions, respectively  ${{{C^q}_q}^{(1)}}$ and  ${{{C^q}_g}^{(1)}}$, Eq.~(\ref{eq:hadfact}), are
      shown.
      Results are shown in the \msbar{}, POS and MPOS schemes. The POS
      scheme is defined in Eqs.~(\ref{eq:counthqq}-\ref{eq:counthqg})
      and (\ref{eq:counthgg}-\ref{eq:counthgq}), and the MPOS scheme
      in Eqs.~(\ref{eq:mposqq}-\ref{eq:mposgg}).
 ${{{C^q}_q}^{(1)}}$ \msbar{} and POS coincide, and the two curves
      correspond, from top to bottom, to \msbar{} and MPOS; for  
      ${{{C^q}_g}^{(1)}}$ POS and MPOS coincide and the two curves
      correspond, from top to bottom, to \msbar{} and POS. 
    \label{fig:dy} }
  \end{center}
\end{figure}
\subsubsection{The diagonal coefficient function}
\label{sec:diag}
We now turn to the diagonal coefficient function: in the \msbar{} scheme
it is given by
\begin{align}\label{eq:cq}
  C_q^{\mmsbar}(z)&=\delta(1-z)+ \frac{\alpha_s}{2\pi}
  {C^{(1)}_q}^{\mmsbar}(z)\\
    &=\delta(1-z) \left(1+\frac{\alpha_s}{2\pi} {\Delta^{(1)}_q}^{\mmsbar}\right)
  +\frac{\alpha_s}{2\pi}  {\overline{C_{q}}^{(1)}}^{\mmsbar}(z) \,,\label{eq:cqnodist}
\end{align}
where in the last step we have separated off the contribution to
$ {C^{(1)}_q}^{\mmsbar}(z)$ proportional to a Dirac $\delta$
(corresponding to a constant in Mellin space) so that
${\overline{C_{q}}^{(1)}}^{\mmsbar}(z)$ only contains  functions and
$+$ distributions.
The NLO diagonal
coefficient function is given by
\begin{align}
{C^{(1)}_q}^{\mmsbar}(z)&=
\lim_{\epsilon\to
  0^-}\left[C_{q}^{(1)}(z,Q^2,\epsilon) - \left(\frac{Q^2}{4\pi
      \mu^2}\right)^{-\epsilon} \left(-\frac 1
    {\epsilon}+\gamma_E\right) P_{qq}(z)\right]\label{eq:cqren}\\
&=\lim_{\epsilon\to0^-}\left[ {C_q^{(1)}}(z,Q^2,\epsilon)
    -\delta_{qq}^{\mmsbar}(z,Q^2,\epsilon)\right]\label{eq:renormq}\\
&= C_F\left[ \left(p_{qq}(z)
    \ln\left(\frac{1-z}{z}\right) \right)_+ - \frac {3}{2}\pDa{z} + 3
    + 2 z - 4 \delta(1-z)\right]\label{eq:cqnlo},
\end{align}
where 
$p_{qq}(z)$ is implicitly defined in terms of the quark-quark
splitting function $P_{qq}(z)$ as
\begin{equation}\label{eq:pqq}
  P_{qq}(z)= C_F\left(p_{qq}(z)\right)_+.
\end{equation}

The Mellin transform of $C^{(1)}_q(z)$ is shown in
Fig.~\ref{fig:dis}. It is clear that the coefficient function is
positive for all $N$: the slightly negative dip of the NLO term
in the $N\sim 1$
region is more than compensated by the much larger LO contribution,
which in $N$ space is a constant (at $\frac{2 \pi}{\alpha_s}$ on the
scale of Fig.~\ref{fig:dis}). 
As $N\to\infty$, where the NLO contribution diverges (and in
principle needs resummation) the growth is actually positive.

A comparison of Eq.~(\ref{eq:cqnlo}) with its off-diagonal counterpart,
Eq.~(\ref{eq:cgnlo}), immediately shows what is going on. In this
case too, the \msbar{} subtraction amounts to an over-subtraction, and
indeed the term proportional to $p_{qq}(z)$ in the coefficient
function Eq.~(\ref{eq:cqnlo}) has the same origin as the term
Eq.~(\ref{eq:logexp}), namely, the collinear singularity due to real
emission, in this case of a gluon from the incoming quark line. In
fact,  this is the contribution which is included in standard
leading-log threshold resummation. Amusingly, the further
(process-dependent)
term, proportional
to $\pDa{z}$, arises at the next-to-leading log level due to collinear
radiation from the outgoing quark line~\cite{Catani:1989ne}, and thus has
the same kinematic origin~\cite{Forte:2002ni}.
One may thus think of generally including these contributions in the
PDF by changing the collinear subtraction, as we did above: indeed
this is done in the Monte Carlo scheme of
Ref.~\cite{Jadach:2016acv}, which aims at including in PDFs
all contributions coming from soft radiation.

However, if the goal is ensure positivity, in the diagonal case
it is not necessary to
modify the \msbar{}  subtraction prescription. Indeed,
in this case over-subtraction actually leads to a more positive
coefficient function, due to the fact that the $P_{qq}$ splitting
function  is negative at large $z$, where it reduces to a $+$
distribution,
i.e., it leads to a negative answer
when folded with a positive test function. Of course, this follows from
baryon number conservation which requires the vanishing of the first
moment of the splitting function. It is in fact easy to check that the
\msbar{} coefficient function, Eq.~(\ref{eq:cq}), is positive for all
$z<1$. The  term proportional to a $\delta$ of course has a positive
coefficient in the perturbative regime, where it is dominated by the
LO term.

We conclude that in order to ensure positivity of the coefficient
function it is sufficient to modify the collinear subtraction only in the
off-diagonal channel. We therefore set
\begin{align}\label{eq:countq}
   {C_{q}^{(1)}}^{\rm DPOS}(z)=  {C_{q}^{(1)}}^{\mmsbar}(z) \,.
\end{align}
Equations~(\ref{eq:count},\ref{eq:countq}) thus define the DPOS
factorization scheme in the quark channel, in terms of the \msbar{} scheme.
Note that the considerations underlying the construction of this
factorization scheme are based on the
structure of the collinear subtraction and the behavior of the
splitting functions, and are therefore process-independent.

In order to fully characterize the scheme it is necessary to also
consider gluon-induced processes. In Ref.~\cite{Altarelli:1998gn},
this was done by considering Higgs production in gluon fusion, with
one of the two gluons coming from a proton and the other being taken
as a pointlike probe. Equivalently, one might consider Higgs production in
photon-gluon fusion. However, the treatment of these
processes is essentially the same as that of hadronic processes, to
which we thus turn.

\subsection{Hadronic processes}
\label{sec:hadr}
For hadronic processes the basic factorization formula has
the same structure as Eq.~(\ref{eq:f2}), with the structure function
replaced by a cross section and the PDF replaced by a parton
luminosity ${\cal L}_{ij}$: up to NLO
\begin{equation}
 \frac{1}{x} \sigma(x,Q^2)= \hat \sigma_0 \left[  {\cal L}_{ii}
   +\frac{\alpha_s}{2\pi}\left( {{C^i}_q}^{(1)} \otimes {\cal L}_{iq} +
 {{C^i}_g}^{(1)} \otimes {\cal L}_{ig}\right) \right],\label{eq:hadfact}
\end{equation}
where for simplicity we consider process for which at LO only one
partonic channel contributes, so
$i=q,g$ labels quark-induced processes (such as Drell--Yan) or
gluon-induced processes (such as Higgs production in gluon fusion),
$\hat{\sigma}_0$ is the LO partonic cross section and
the parton luminosity is 
\begin{equation}\label{eq:lumi}
 {\cal L}_{ij}= f_i\otimes f_j\,.
\end{equation}

We first discuss quark-induced processes: their treatment is very
close to that of DIS  presented in the previous
section, so it is sufficient to highlight the differences. We then 
turn to gluon-induced processes, for which we repeat the analysis of Sect.~\ref{sec:discf}.

\begin{figure}[t]
  \begin{center}
    \includegraphics[width=0.8\linewidth]{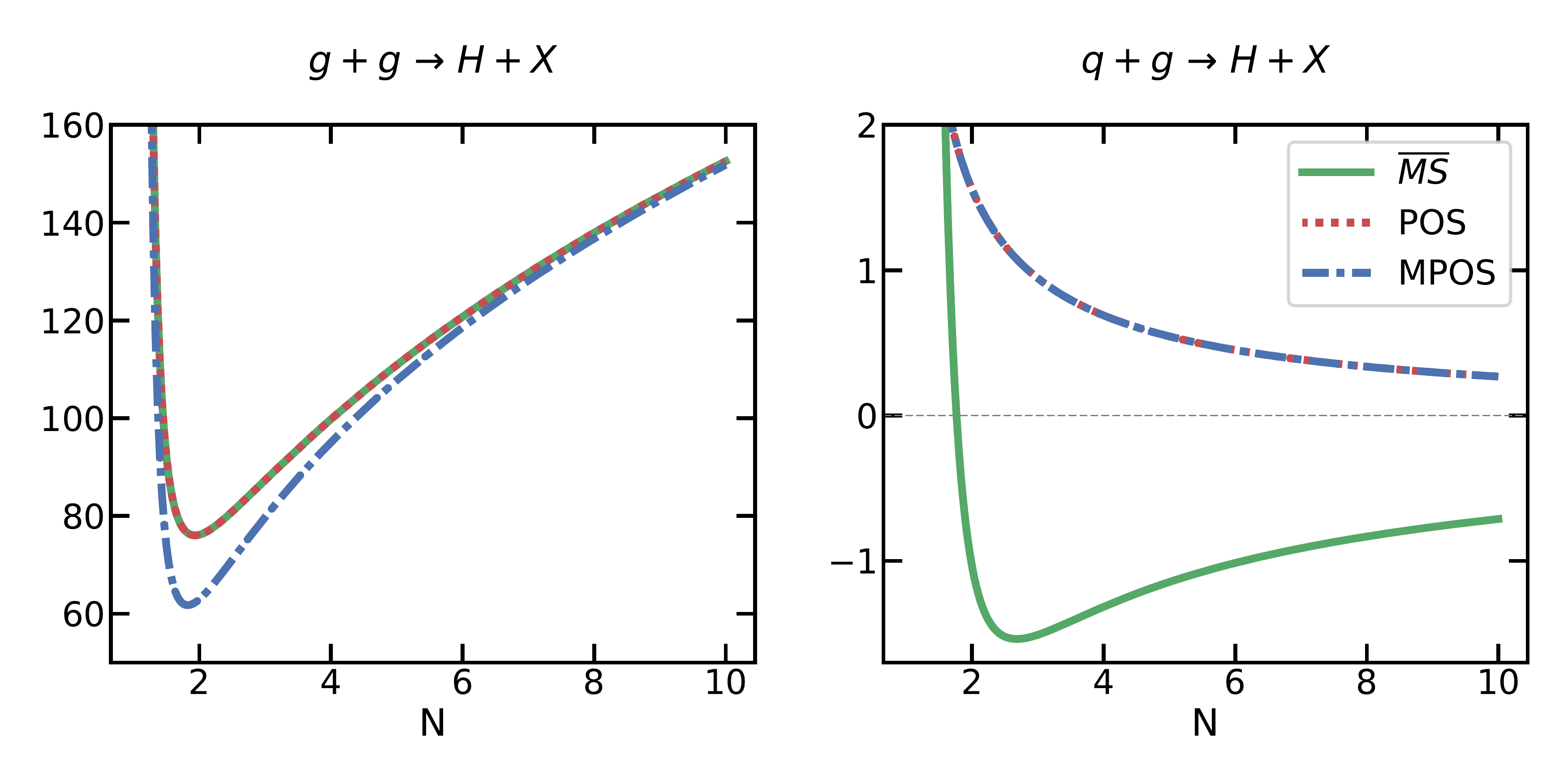}
    \caption{\small Same as Fig.~\ref{fig:dy}, but now for the Higgs
      coefficient functions  ${{{C^g}_g}^{(1)}}$ (left) and
      ${{{C^g}_q}^{(1)}}$ (right).
    \label{fig:higgs} }
  \end{center}
\end{figure}

\subsubsection{Quark-induced processes}
\label{sec:dy}
As a prototype of quark-induced process we consider Drell--Yan
production. The NLO coefficient functions (i.e.\ NLO partonic
cross sections normalized to the LO result) are given by 
\begin{align}\label{eq:dyq}
 {{C^q{}_q}^{(1)}}^{\mmsbar}(z)&= C_F\left[
 \left(\frac{4\pi^2}{3} - \frac 7 2\right)\delta(1-z) + 2
 \left(p_{qq}(z) \ln\left(\frac{(1-z)^2}{z}\right)
 \right)_+\right]\nonumber  \\
 &={\Delta_{qq}^{(1)}}^{\mmsbar}\delta(1-z) + 2 C_F \left(p_{qq}(z) \ln\left(\frac{(1-z)^2}{z}\right) \right)_+,\\
 \label{eq:dyg}
  {{C^q{}_g}^{(1)}}^{\mmsbar}(z)&= P_{qg}(z) \left[ \ln\left(\frac{(1-z)^2}{z}\right) - 1 \right]
  + C_F\left[\frac 3 2 -  \frac 3 2 z^2 + z\right].
\end{align}

Comparing the coefficient functions
Eq.~(\ref{eq:dyq}-\ref{eq:dyg}) to their DIS counterparts
Eqs.~(\ref{eq:cgren},\ref{eq:cqnlo}) shows that they have the same
structure, with a residual logarithmic contribution proportional to
the splitting function, due to over-subtraction. The only difference is
that the argument of the log is now $\frac{(1-z)^2}{z}$. This is again
recognized to be the upper limit of the transverse momentum integral,
and to coincide with the argument of the logs whose
renormalization-group improvement leads to threshold
resummation~\cite{Forte:2002ni}: indeed,
for a $2\to2$ process with a final state particle with mass $M^2$, and
$z=\frac{M^2}{s}$,
\begin{equation}\label{eq:ktmaxh}
\mu_h^2=  |k_T^{\rm max,\, had}|^2=\frac{(s-Q^2)^2}{4s}=\frac{Q^2(1-z)^2}{4z}\,,
\end{equation}
where $ Q^2 = M^2 $.
The coefficient
functions, Eq.~(\ref{eq:dyq}-\ref{eq:dyg}), are displayed in
Fig.~\ref{fig:dy} in Mellin space; their qualitative features are the
same as those of the DIS coefficient functions.

Hence, just as in case of DIS, it is possible to define a positive
subtraction scheme, which we call POS, and which differs from \msbar{}
because in the off-diagonal quark-gluon channel the
subtraction is performed at the scale $\mu_h^2$,
Eq.~(\ref{eq:ktmaxh}). Just like for DIS, in the diagonal quark-quark
channel there is no need to modify the \msbar{} subtraction, which
actually makes the coefficient function more positive, so we define 
a POS factorization of the DY process according to
\begin{align}\label{eq:counthqq}
 {{C^q{}_q}^{(1)}}^{\rm POS}(z) &=  {{C^q{}_q}^{(1)}}^{\mmsbar}(z) \,, \\
 {{C^q{}_g}^{(1)}}^{\rm POS}(z) &=  {{C^q{}_g}^{(1)}}^{\mmsbar}(z) - K_{qg}^{\rm POS} (z) \,,\\ \label{eq:counthqg}
  K_{qg}^{\rm POS} (z) &=  P_{qg}(z)\left[\ln\left(\frac{(1-z)^2}{z}\right) - 1\right] \,.
\end{align}
The quark-gluon coefficient function can be read off
Eqs.~(\ref{eq:dyg},\ref{eq:counthqg}) and it is easy to check that it
is positive definite for all $0<z<1$.

Of course, a choice of factorization scheme must be
universal. Therefore, it is interesting to check what this choice
amounts to if adopted for DIS. Clearly, the
hadronic scale Eq.~(\ref{eq:ktmaxh}) is always lower than the DIS
scale Eq.~(\ref{eq:ktmaxh}): $\mu_h^2< \mu_D^2$. Hence, subtraction in the
DPOS scheme amounts to under-subtraction, and if adopted for DIS
coefficient function it leads to a DIS coefficient function
${C^{(1)}_{g}}^{\rm POS}(z)$ which is actually more positive than that
in the DPOS scheme. This is seen in Fig.~\ref{fig:dis} (right), where 
${C^{(1)}_{g}}(z)$ is shown in the \msbar{}, DPOS and POS schemes.

\subsubsection{Gluon-induced processes}
\label{sec:higgs}
In order to fix completely the factorization scheme we turn to
gluon-induced hadronic processes. We choose Higgs production in gluon
fusion (in the infinite top mass limit) as a prototype, and we repeat the  analysis of
Sect.~\ref{sec:offdiag}, but now for the quark coefficient function
${{{C^g}_q}^{(1)}}$. The regularized, unsubtracted expression is (see
e.g.~\cite{Maltoni:2018dar}) 
\begin{equation}\label{eq:cgqeps}
{C^g{}_q}^{(1)}(z,Q^2,\epsilon) =
  \frac{ \Gamma(-\epsilon)
  \left(\frac{\mu_h^2}{\pi\mu^2}\right)^{-\epsilon} (1-\epsilon) \left[P_{gq}(z) - C_F \frac{(1+z)^2}{2z} \epsilon \right]}{16\pi (2 - 2\epsilon) \Gamma (3 - 2 \epsilon)}\,,
\end{equation}
where $\mu_h^2$ is given by Eq.~(\ref{eq:ktmaxh}), with $Q^2=M_H^2$,
the Higgs square mass.
Performing  \msbar{} subtraction in the usual way we get
\begin{align}\label{eq:cgqren}
{  {C^g{}_q}^{(1)}}^{\mmsbar}(z) &= \lim_{\epsilon\to
  0^-}\left[{{C^{g}}_q}^{(1)}(z,Q^2,\epsilon) - \left(\frac{Q^2}{4\pi
      \mu^2}\right)^{-\epsilon} \left(-\frac 1
    {\epsilon}+\gamma_E\right) P_{gq}(z)\right]\\
&= P_{gq}(z)\left[\ln\left(\frac{(1-z)^2}{z}\right) - 1 \right]  + C_F \frac{(1+z)^2}{2z} 
 \,.\label{eq:cgqnlo}
\end{align}

Again, we encounter the same situation that we have seen in the quark
channel for DIS, Eqs.~(\ref{eq:cgeps},\ref{eq:cgren}): the collinear
log has a scale set by the upper limit of the transverse momentum
integration, now the hadronic $\mu_h^2$, Eq.~(\ref{eq:ktmaxh}), but the
\msbar{} subtraction is performed at the scale $Q^2$, which at large $z$
is higher, thus leading to over-subtraction. Indeed, the Mellin-space
\msbar{} coefficient function  ${{{C^g}_q}^{(1)}}$, shown in
Fig.~\ref{fig:higgs}, is seen to be negative at large $N$.

As in the quark sector, the problem is fixed by performing the
collinear subtraction at the physical scale $\mu_h^2$. Note that
also in this case, as for the DIS quark-gluon channel,
there is an issue with the sum over gluon polarizations: indeed,
because the LO process is in the gluon-gluon channel, even the NLO
quark-gluon channel has a gluon in the initial state, leading
to a factor  $1-\epsilon$ in the 
denominator of Eq.~(\ref{eq:cgqeps}), which must be accounted for in
order to avoid over-subtraction.
Hence, we define the POS scheme coefficient function as
\begin{align}\label{eq:cqrenposh}
  {{C^g{}_q}^{(1)}}^{\rm POS} (z)  &= \lim_{\epsilon\to0^-}
  \left[  {C^g{}_q}^{(1)}(z,Q^2,\epsilon) - \frac{1}{1-\epsilon}\left(\frac{\mu_h^2}{\pi\mu^2}\right)^{-\epsilon} \left(-\frac 1
    {\epsilon}+\gamma_E\right) P_{qg}(z)\right]\\ \label{eq:cqrenposh1}
    &=C_F \frac{(1+z)^2}{2z} \, ,
\end{align}
with $\mu_h^2$ given by Eq.~(\ref{eq:ktmaxh}).
The coefficient function is clearly positive. Its Mellin transform is
also shown in Fig.~\ref{fig:higgs}.

We finally examine the gluon-gluon NLO coefficient function:
\begin{align}\label{eq:higgsg}
\hspace*{-30pt}  {{C^g{}_g}^{(1)}}^{\mmsbar} (z) &=C_A\left[ 2\frac 1 z \left(z p_{gg}(z) \ln\left(\frac{(1-z)^2}{z}\right) \right)_++
    \left(\frac{473}{36} + \frac{4\pi^2}{3}\right)\delta(1-z) -
    \frac{11}{3}\frac{(1-z)^3}{z}\right]\nonumber\\
&={\Delta^{(1)}_{gg}}^{\mmsbar}\delta(1-z)+ C_A\left[ 2\frac 1 z \left(z p_{gg}(z) \ln\left(\frac{(1-z)^2}{z}\right) \right)_+-
    \frac{11}{3}\frac{(1-z)^3}{z}\right] \,,
\end{align}
where, in analogy to Eq.~(\ref{eq:pqq}), $p_{gg}(x)$ is implicitly
defined by
\begin{equation}\label{eq:pgg}
  P_{gg}(z)= C_A\frac 1 z \left(z p_{gg}(z)\right)_+ - \frac{n_f}{3} \delta(1-z).
\end{equation}
As in the diagonal quark channel, the \msbar{} subtraction is now
multiplied by a splitting function which is negative at large $z$, for
the same physical reason. It therefore leads to a coefficient function
which is positive, as seen by inspecting Eq.~(\ref{eq:higgsg}) and
shown in Fig.~\ref{fig:higgs} (left), so no further scheme change is needed.

Therefore we get
\begin{align}\label{eq:counthgg}
 {{C^g{}_g}^{(1)}}^{\rm POS}(z) &=  {{C^g{}_g}^{(1)}}^{\mmsbar}(z) \,, \\
 {{C^g{}_q}^{(1)}}^{\rm POS}(z) &=  {{C^q{}_g}^{(1)}}^{\mmsbar}(z) - K_{gq}^{\rm POS} (z) \,,\\ \label{eq:counthgq}
  K_{gq}^{\rm POS} (z) &=  P_{gq}(z)\left[\ln\left(\frac{(1-z)^2}{z}\right)
    - 1\right].
\end{align}

Equations~(\ref{eq:counthqq}-\ref{eq:counthqg}) and (\ref{eq:counthgg}-\ref{eq:counthgq}) 
fully define the POS subtraction. We shall see in the next section
that they define a positive factorization scheme. Indeed, in the
construction presented in this section we have not made use of the
detailed from of the partonic cross section, but rather just of the
collinear counterterms, expressed in terms of universal splitting
functions. Hence, these counterterms, when used in
Eq.~(\ref{eq:pertz})  define a universal renormalization scheme
Eq.~(\ref{eq:pdffac}) for PDFs, without spoiling PDF universality.

\section{A positive factorization scheme}
\label{sec:scheme}

We will now construct a positive factorization scheme based on the POS 
subtraction of
Eqs.~(\ref{eq:counthqq}-\ref{eq:counthqg},\ref{eq:counthgg}-\ref{eq:counthgq}).
We then discuss the scheme transformation from this scheme to the
\msbar{} scheme and use it to show that PDFs are non-negative in the
\msbar{} scheme in the perturbative region.

The argument is based on the factorization
Eqs.~(\ref{eq:refaca}-\ref{eq:refacb}), and, very crudely speaking,
amounts to showing that with the POS subtraction, all factors in
Eqs.~(\ref{eq:refacb}) are positive: the left-hand side is positive
because it is a physically measurable cross-section, the coefficient $C^S$
function on the right-hand side is positive because the POS
subtraction preserves the positivity of the unsubtracted coefficient
function $C$, which is a partonic cross-section, and thus positive
before subtraction, but only well-defined in $d>4$
dimensions.

Taking a Mellin transform of both sides of
Eqs.~(\ref{eq:refaca}-\ref{eq:refacb}) all convolutions turn into
ordinary products, and it is immediately clear that, because the left-hand
side is positive, for the
Mellin transformed PDF to be positive it is necessary and sufficient
that the coefficient  function is positive. However, 
positivity of the Mellin transform of a function is a necessary
condition for its positivity, but not a sufficient one: a negative
function may have a positive Mellin transform. The somewhat more
complex structure of the discussion below is necessary in order
to deal with the necessity of providing an $x$-space argument.

\subsection{Positive PDFs}
\label{sec:pospdf}

We start by presenting the construction in a simplified setting, 
namely in the absence of parton mixing. This means that the operators
Eq.~(\ref{eq:quark}) whose matrix elements define the PDFs renormalize
multiplicatively.
This would specifically correspond to
the case of a quark combination that does not mix with the
gluon, such as  any combination $q^{\NS}(Q^2)=q_i(Q^2)-q_j(Q^2)$,
where $i,\>j$ denote generically a quark flavor or antiflavor, with
$i\not= j$.  We refer to this as a 
nonsinglet quark combination. We can think of the argument below as
applying to such a combination, 
chosen in such a way that  the bare
$q^{\NS}(Q^2)^{(0)}$, Eq.~(\ref{eq:quark}), is positive --- which in
general of course will not be true even if $q_i$ and $q_j$ are
separately positive. 
This should be viewed as an academic case --- after
all, in principle, a positive nonsinglet PDF might not exist --- whose
purpose is to illustrate the structure of the argument in the absence
of parton mixing. We then turn
to the realistic case of PDFs that do undergo mixing upon
renormalization (which we will refer to as singlet case).
The nonsinglet case is simpler, not only because of the absence of
mixing, but also because in  this case
the POS scheme actually coincides with \msbar{} (i.e., \msbar{} is already
positive). 

\subsubsection{The nonsinglet case as a toy model}
\label{sec:nonsing}

In the nonsinglet case, only the diagonal
quark subtraction is relevant: so in the nonsinglet case
the DIS structure function Eq.~(\ref{eq:f2}) becomes
\begin{equation}\label{eq:disfact}
 \frac{1}{x} F_2^{\NS}(x,Q^2)= \langle e_i^2\rangle \left[1
 +\frac{\alpha_s}{2\pi}  C^{(1)}_q \otimes \right] q^{\NS}(Q^2) \,,
\end{equation}
where $q^{\NS}$ is a difference of two quark or antiquark PDFs, assumed
positive and
$\langle
e^2_i\rangle=\frac{1}{2} \left(e^2_i+e^2_j\right)$ is the average of
their electric charges. 

The factorization Eqs.~(\ref{eq:refaca}-\ref{eq:refacb}) takes the
form
\begin{align}
  \frac{1}{x} F_2^{\NS}(x,Q^2)&= \langle e_i^2\rangle
\lim_{\epsilon\to
  0^-}\left[1
    +\frac{\alpha_s}{2\pi} C_{q}^{(1)}(Q^2,\epsilon) \otimes \right] 
    \left[{q^{\NS}}\right]^{(0)} \label{eq:dismsbar} \\
&= \langle e_i^2\rangle
\lim_{\epsilon\to
  0^-}\left[1
  +\frac{\alpha_s}{2\pi} {C_{q}^{(1)}}^{\mmsbar}(Q^2, \epsilon) \otimes \right] 
\left[1
    +\frac{\alpha_s}{2\pi} \delta^{\mmsbar} (Q^2,\epsilon) \otimes \right]
     \left[{q^{\NS}}\right]^{(0)} \label{eq:dismsbar1} \\
    &=  \langle e_i^2\rangle
\left[1+ \frac{\alpha_s}{2\pi} {\Delta^{(1)}_q}^{\mmsbar}
  +\frac{\alpha_s}{2\pi}    {\bar{C}_{q}^{(1)}{}}^{\mmsbar} \otimes
  \right] 
     \left[{q^{\NS}}\right]^{\mmsbar}(Q^2) \,, \label{eq:dismsbar2}
\end{align}
where $ {C_{q}^{(1)}}^{\mmsbar}$,  ${\overline{C_{q}}^{(1)}}^{\mmsbar}$,
${\Delta^{(1)}_q}^{\mmsbar}$ and $ \delta_{qq}^{\mmsbar} $ have been
defined in Eqs.~(\ref{eq:cqnodist},\ref{eq:cqren},\ref{eq:renormq}),
and the dependence on $x$ on the  right-hand side has been omitted
because it appears due to the convolution, while the dependence on all
other variables has been indicated explicitly.

Now, the discussion of Sect.~\ref{sec:discf} shows that, because the
bare PDF of Eq.~(\ref{eq:quark}) is a probability density, the three
factors which are convoluted in Eq.~(\ref{eq:dismsbar2}) are all
separately positive when $\epsilon \to 0^-$, i.e.\ from the negative
region, provided only $\mu^2< \mu_D^2$, with $\mu_D^2$ given by
Eq.~(\ref{eq:mud})\footnote{Note that the condition cannot be satisfied in
the strict $x\to1$ limit, but this is as it should be since in the
limit the scattering process becomes elastic and it is no longer
described by perturbative QCD.}.
This, as discussed in Sect.~\ref{sec:diag} [see in particular
Eq.~(\ref{eq:cqnlo}) and Fig.~\ref{fig:dis}]
can be understood as a consequence of the fact that the only region in
which the $O(\alpha_s)$ term could overwhelm the LO contribution is
the threshold region $z\to 1$, where $\alpha_s\ln(1-z)\sim1$.
However, in this region the \msbar{} over-subtraction leads to a
coefficient function which is positive because $P_{qq}$ is negative at
large $z$. Consequently, all factors in  Eq.~(\ref{eq:dismsbar2})
remain positive for all $z$.

The  meaning of the factorization argument
Eqs.~(\ref{eq:dismsbar}-\ref{eq:dismsbar2}) can be understood by
noting
that it is possible to choose a ``physical'' factorization
scheme~\cite{Catani:1995ze} in which PDFs are identified with physical
observables. This means that the coefficient function is set to one to
all orders by scheme choice. An example is the ``DIS'' scheme~\cite{Diemoz:1987xu} in which
the quark PDF is identified with the DIS structure function, so that
Eq.~\ref{eq:disfact} becomes
\begin{equation}\label{eq:disfactns}
 \frac{1}{x} F_2^{\NS}(x,Q^2)= \langle e_i^2\rangle  \left[{q^{\NS}}\right]^{\rm DIS}(x,Q^2) \,,
 \end{equation}
which holds to all perturbative orders. Comparing this DIS scheme
expression of the structure function to the \msbar{} expression,
Eq.~(\ref{eq:f2}), immediately shows that the quark PDF in the DIS
and \msbar{} schemes are related by
\begin{equation}\label{eq:msbartodis}
\left[{q^{\NS}}\right]^{\rm DIS}(\xi,Q^2)=
\left[1+\frac{\alpha_s}{2\pi} {\Delta^{(1)}_q}^{\mmsbar}
  +\frac{\alpha_s}{2\pi}
  {{{\bar{C}}^{(1)}}_q{}}^{\mmsbar}\otimes\right]\left[{q^{\NS}}\right]^{\mmsbar}(Q^2) \,,
\end{equation}
where again we have dropped the $\xi$ dependence of the convolution
on the right-hand side,
as in Eqs.~(\ref{eq:dismsbar}-\ref{eq:dismsbar2}).

The \msbar{} PDFs can be obtained in terms of the DIS ones by inverting
Eq.~(\ref{eq:msbartodis}): perturbative inversion of course gives
\begin{equation}\label{eq:distomsbar}
\left[{q^{\NS}}\right]^{\mmsbar}(\xi,Q^2)=
\left[1-\frac{\alpha_s}{2\pi} {\Delta^{(1)}_q}^{\mmsbar}
  -\frac{\alpha_s}{2\pi}  {\bar
    C^{(1)}_q{}}^{\mmsbar}\otimes\right]\left[{q^{\NS}}\right]^{\rm DIS}(Q^2)
+O(\alpha_s^2) \,.
\end{equation}
One may worry that therefore the \msbar{} PDFs may turn negative in the
large $\xi$ region, where $\alpha_s\ln(1-\xi)\gtrsim 1$ and the last term
in square brackets in Eq.~(\ref{eq:distomsbar}), which is negative, may overwhelm the LO
contribution 
term. However, in this region the perturbative
inversion is invalid, but it is easy to invert
Eq.~(\ref{eq:msbartodis}) exactly in the asymptotic large $\xi$ limit.
Letting
\begin{align}\label{eq:msbartodisp}
    \hspace{-20pt}\left[{q^{\NS}}\right]^{\rm DIS}(\xi,Q^2)&=
\left[1+\frac{\alpha_s}{2\pi} {\Delta^{(1)}_q}^{\mmsbar}
  +\frac{\alpha_s}{2\pi} 2C_F\left[\frac{\ln(1-z)}{1-z}\right]_+
  \otimes\right]\left[{q^{\NS}}\right]^{\mmsbar}(Q^2)\nonumber\\
    &\qquad+ \hbox{NLL}(1-\xi)\\
\label{eq:msbartodisp1}
    &= \left(1+\frac{\alpha_s}{2\pi} {\Delta^{(1)}_q}^{\mmsbar}\right) 
\left[1+c_{\rm LL}\left[\frac{\ln(1-z)}{1-z}\right]_+
\otimes\right]\left[{q^{\NS}}\right]^{\mmsbar}(Q^2)\nonumber\\
    &\qquad+ \hbox{NLL}(1-\xi) \,,
\end{align}
with
\begin{equation}
    \label{eq:msbartodiscoeff}
    c_{\rm LL} = \frac{\frac{\alpha_s}{2\pi} 2C_F}{1+\frac{\alpha_s}{2\pi} {\Delta^{(1)}_q}^{\mmsbar}} \,, 
\end{equation}
and which holds at the leading $\ln(1-\xi)$ level (LL$(1-\xi)$),
inversion can be performed by going to Mellin space and then computing
the Mellin inverse term by term in an expansion in powers of $\alpha_s$.
We get
\begin{align}\label{eq:distomsbarllx}
    \hspace*{-20pt} \left[{q^{\NS}}\right]^{\mmsbar}(\xi,Q^2) &= \frac{1}{1+\frac{\alpha_s}{2\pi} {\Delta^{(1)}_q}^{\mmsbar}} \times \nonumber\\
    &\hspace*{-60pt}\Bigg[1 -c_{\rm LL} \left[ \frac{\ln(1-z)}{\left[1+c_{\rm LL}\ln^2(1-z)/2\right]^2}  \frac{1}{1-z}\right]_+
\otimes\Bigg]\left[{q^{\NS}}\right]^{\rm DIS}(Q^2) + \hbox{NLL}(1-\xi) \,.
\end{align}
It is clear that as $\xi\to 1$ the negative LL$(1-\xi)$ contribution
actually vanishes.\footnote{A similar argument also applies at small
  $\xi$, where the coefficient function also rises, as seen in
  Fig.~\ref{fig:dis}. We do not discuss this case in detail since
  positivity of the \msbar{} PDF at small $\xi$ is manifest.}

Now, we observe that
$\left[{q^{\NS}}\right]^{\rm DIS}(\xi,Q^2)$ is positive because it is a physical
observable. Equation~(\ref{eq:msbartodis}), which expresses the DIS
PDF in terms of the \msbar{} one, then  implies that  for
$\left[{q^{\NS}}\right]^{\mmsbar}(\xi,Q^2)$  to be guaranteed to be
positive,
the \msbar{} coefficient function  must also be positive, otherwise
folding a positive \msbar{}  PDF with a negative coefficient function
could lead to a negative DIS PDF. So positivity of the \msbar{}
coefficient function
is a necessary condition for positivity of the \msbar{} PDF.
However, the inverse of  Eq.~(\ref{eq:msbartodis}), expressing the
\msbar{} PDF in terms of the DIS one, implies that the condition is
also sufficient, because it gives the \msbar{} PDF as the convolution
of a positive coefficient with a positive PDF. Equations.~(\ref{eq:distomsbar},\ref{eq:distomsbarllx}) show that the coefficient is indeed
positive because in the dangerous
$\xi\to 1$ region, where  a large negative contribution may arise,
inversion can be performed exactly and shown to lead to a positive
result.
Of
course, this argument works for any factorization scheme, and it shows
that a necessary and (perturbatively) sufficient condition for the
PDFs to be positive is that the coefficient function in that scheme is positive.

The perturbative nature of the argument is worth commenting
upon. As discussed at the beginning of this section, the corresponding
Mellin space argument is trivial: because in Mellin space the
structure function is the product of the PDF times the coefficient
function, it follows that positivity of the coefficient function is
necessary and sufficient for the positivity of the PDF. However, as
already mentioned,
Mellin-space positivity is not sufficient for $x$-space positivity. It
is therefore necessary to compute the $x$-space inverse of the
coefficient function, and check that it is still positive.

The
inversion is done perturbatively in Eq.~(\ref{eq:distomsbar}), and it
leads to a coefficient function which is manifestly positive in most
of the $z$ range, except at small and large $z$, where the coefficient
functions blows up, due to high-energy (BFKL) and soft (Sudakov) logs
respectively. Consider the large-$z$ case that was discussed
above. Upon 
Mellin transformation, the $z\to 1$ region is mapped onto the
$N\to\infty$ region, and specifically, as well known (see
e.g. Ref.~\cite{Forte:2002ni}) powers of $\ln(1-z)$ are mapped onto
powers of $\ln N$.  The $\ln N$
  logarithmic growth of the coefficient function in this limit is seen in Fig.~\ref{fig:dis}, where it
  is apparent that the coefficient function diverges as
  $N\to\infty$. The $N$-space inverse of the coefficient function is
  just its reciprocal, and thus it manifestly vanishes as
  $N\to\infty$ (while of course remaining positive). One would
  therefore naively expect that the $x$-space inverse also vanishes
  (from the positive side) as $z\to1$, and this expectation is borne
  out by the explicit computation presented above in
  Eq.~(\ref{eq:distomsbarllx}).\footnote{In view of the fact that the
    Mellin space inverse coefficient function behaves as $[\bar
      C^{(1)}(N)]^{-1}\toinf{N}\frac{1}{\ln^2 N}$ it may appear surprising hat
    the term in square brackets in Eq~(\ref{eq:distomsbarllx}) starts
    with one. However, it should be born in mind that the Mellin
    transform of any function which is regular (or indeed integrable)
    at $x=1$ vanishes as $\frac{1}{N^k}$, with $k>0$, hence in Mellin space the
    suppression of the inverse coefficient function as $N\to\infty$ is
    a subleading correction to the leading power suppression of
    $q^{\rm NS}(N)$.}  Similar arguments apply at higher orders (NNLO
  and beyond), where the coefficient function grows with a higher
  order power of $\ln(1-z)$ as $z\to1$, and at small $z$, where the
  coefficient function grows as powers of $\ln \frac{1}{z}$ as
    $z\to0$. Hence, either the coefficient function is not
    logarithmically enhanced, and then the perturbative inverse is
    manifestly positive, or it is logarithmically enhanced, and then
    the exact inverse of the enhanced terms can be computed ans also
    shown to be positive. It is natural to conjecture that an explicit
    computation of the exact inverse of the full coefficient function
    would also be positive.

The perturbative  assumption is therefore used in two different
ways. On the one hand, the NLO correction to the \msbar{}  coefficient function
 ${\bar  C^{(1)}_q(z)}^{\mmsbar}$ is not everywhere positive, as it is
apparent from Fig.~\ref{fig:dis}. However, this is a small correction
to the positive coefficient function if $\alpha_s\lesssim 1$, and the
overall coefficient function remain positive. This would fail in a
region in which $\alpha_s$ blows up. So the full NLO coefficient
function remains positive, but only in the perturbative region. On the other
hand, the perturbative inversion Eq.~(\ref{eq:distomsbar}) is used to
show that positivity of the coefficient function is shared by its
inverse, and in regions in which perturbativity would fail it is
checked explicitly that this is the case by exact inversion. In this
case we conjecture that positivity of the inverse is actually an exact
property, even when $\alpha_s$ is arbitrarily large.

The argument based on the physical factorization scheme showing that
a positive coefficient function is necessary and (perturbatively)
sufficient for a positive PDF is in fact equivalent to the
factorization argument
Eqs.~(\ref{eq:dismsbar}-\ref{eq:dismsbar2}). Indeed, 
the operator definition of the quark distribution, Eq.~(\ref{eq:quark}),
upon performing a derivative expansion of the Wilson line, leads to
the standard expression of its moments in terms of matrix elements of
local operators. The interpretation of the bare quark distribution as
a probability  is then
preserved by any physical subtraction scheme such that the matrix
elements of Wilson operators are expressed in terms of a measurable
quantity. The DIS scheme of Eq.~(\ref{eq:disfact}) is of course an
example of this scheme. Given the equivalence of the two arguments,
one may wonder whether, if at all, perturbativity is used in the
argument of Eqs.~(\ref{eq:dismsbar}-\ref{eq:dismsbar2}): specifically,
the perturbative inversion of Eq.~(\ref{eq:distomsbar}). The question is
answered in the affirmative: the perturbative inversion is hidden in the step leading from Eq.~(\ref{eq:dismsbar})
to Eq.~(\ref{eq:dismsbar1}). Indeed, this step amounts to 
\begin{equation}\label{eq:exinv}
\left[1
  +\frac{\alpha_s}{2\pi} {C_{q}^{(1)}}^{\mmsbar}\otimes\right]^{-1} \left[1
    +\frac{\alpha_s}{2\pi} C_{q}^{(1)}(Q^2,\epsilon)\right]=
\left[1
    +\frac{\alpha_s}{2\pi} \delta^{\mmsbar} (Q^2,\epsilon)\right]
+O(\alpha_s^2) \,,
\end{equation}
i.e.\ the perturbative inversion of the \msbar{} coefficient function. 
The two arguments are thus seen
to coincide. Again, while we only provide a perturbative argument it
is natural to conjecture that the argument is in fact exact (i.e.\ it
also holds for large values of $\alpha_s$).

\subsubsection{The POS factorization scheme}
\label{sec:pos}

Equipped with the results of Sect.~\ref{sec:nonsing} we can  turn to
the case in which parton mixing is present.   This corresponds to the
realistic case in which the operators Eq.~(\ref{eq:quark}) mix with
the gluon and conversely (at NLO) and with each other at NNLO and
beyond. Because at NLO only quark-gluon mixing is present, we refer to
this as the singlet case.
In order to fully define
the factorization scheme at NLO we must thus consider a pair of
processes, a quark-induced and a gluon-induced one.
The factorization for a pair of hadronic processes
can be written as 
\begin{equation}\label{eq:singhadfact}
 \frac{1}{x} \sigma(x,Q^2)= \hat \Sigma_0\otimes \left[1
 +\frac{\alpha_s}{2\pi}  C^{(1)} \otimes \right] f(Q^2) \,.
 \end{equation}
In Eq.~(\ref{eq:singhadfact}) \begin{itemize}
  \item $\sigma(x,Q^2)$ is a vector of hadronic cross sections
\begin{equation}\label{eq:hadxsec}
   \sigma(x,Q^2)=\left(\begin{array}{c} \sigma^{q}(x,Q^2)\\ \sigma^{g} (x,Q^2)\end{array}\right),
\end{equation}
 such as
the pair of processes of Sect.~\ref{sec:hadr}, namely Drell--Yan  and
Higgs production in gluon fusion; we are assuming for simplicity  and without loss of
generality that both are evaluated at the same scale $Q^2=M^2$ (such
as when producing an off-shell gauge boson and/or Higgs with the same
mass), with a trivial generalization to the case of unequal scales,
and the scaling variable is $x=\frac{Q^2}{s}$, with $s$ the hadronic
center-of-mass energy;
\item $\hat \Sigma_0$ is a diagonal matrix of LO partonic cross sections,
  multiplied by the respective PDFs, 
\begin{equation}\label{eq:lohadxsec}
   \hat \Sigma_0(x,Q^2)=\left(\begin{array}{cc} \hat \sigma_0^{q} q (x,Q^2) &
   0 \\0 & \sigma_0^{g} g (x,Q^2) \end{array}\right),
\end{equation}
namely the quark and the gluon respectively for Drell--Yan and Higgs;
\item $C^{(1)}$ is the two-by-two matrix of NLO 
  coefficient functions ${C^i{}_{j}}^{(1)}$ with $i,\, j=q,\, g$
  defined in Eq.~(\ref{eq:hadfact});
\item $f(\xi,Q^2)$ is a vector of PDFs that mix upon renormalization:
\begin{equation}\label{eq:singpdf}
   f(\xi,Q^2)=\left(\begin{array}{c} q(\xi,Q^2) \\ g(\xi,Q^2) \end{array}\right).
\end{equation}
\end{itemize}

Having established a suitable notation, the argument then proceeds
in an analogous way as the nonsinglet argument of
Sect.~\ref{sec:nonsing}, except that now, in order to guarantee
positivity of the two-by-two matrix of coefficient functions, we must
perform the POS subtraction, which in the diagonal
channels (and thus in the nonsinglet case) coincides with \msbar{} but
in the off-diagonal channel differs from it.
Namely, we have
\begin{align}
\frac{1}{x} \sigma(x,Q^2)&= \hat \Sigma_0\otimes
\lim_{\epsilon\to
  0^-}\left[\mathbb{I}
    +\frac{\alpha_s}{2\pi} C^{(1)}(Q^2,\epsilon) \otimes\right]
    f^{(0)} \label{eq:posmsbar} \\
&= \hat \Sigma_0\otimes
\lim_{\epsilon\to
  0^-}\left[\mathbb{I}
  +\frac{\alpha_s}{2\pi} {C^{(1)}}^{{\rm POS}}(Q^2,\epsilon) \otimes\right]
\left[\mathbb{I}
    +\frac{\alpha_s}{2\pi} \delta^{{\rm POS}} (Q^2,\epsilon)\otimes\right]
    f^{(0)} \label{eq:posmsbar1} \\
    &=  \hat \Sigma_0\otimes
\left[\mathbb{I}+ \frac{\alpha_s}{2\pi} {\Delta^{(1)}}^{{\rm POS}}
  +\frac{\alpha_s}{2\pi}    {\overline{C}^{(1)}}^{{\rm POS}} \otimes
  \right] 
     f^{{\rm POS}}(Q^2) \,. \label{eq:posmsbar2}
\end{align}
In Eqs.~(\ref{eq:posmsbar}-\ref{eq:posmsbar2})\begin{itemize}
\item $ {\Delta^{(1)}}^{{\rm POS}}$ is the diagonal matrix
\begin{equation}\label{eq:Deltamat}
  {\Delta^{(1)}}^{{\rm POS}}=\left(\begin{array}{cc} {\Delta_{qq}^{(1)}}^{\mmsbar} &
   0 \\0 & {\Delta_{gg}^{(1)}}^{\mmsbar} \end{array}\right),
\end{equation}
with  ${\Delta_{ii}^{(1)}}^{\mmsbar}$ defined in Eqs.~(\ref{eq:dyq}) and
(\ref{eq:higgsg}) respectively for $i=q$ and $i=g$;
\item $\delta^{{\rm POS}} (Q^2,\epsilon)$ is a two-by-two matrix of
  counterterms
\begin{equation}\label{eq:deltamat}
 \delta^{{\rm POS}} (z,Q^2,\epsilon)= \left(-\frac 1
    {\epsilon}+\gamma_E\right) \left(\begin{array}{cc}  \left(\frac{Q^2}{4\pi
      \mu^2}\right)^{-\epsilon} P_{qq}(z) & \frac{1}{1-\epsilon}\left(\frac{\mu_h^2}{\pi\mu^2}\right)^{-\epsilon} 
  P_{qg}(z)   \\\frac{1}{1-\epsilon}\left(\frac{\mu_h^2}{\pi\mu^2}\right)^{-\epsilon} P_{gq}(z)  & \left(\frac{Q^2}{4\pi
      \mu^2}\right)^{-\epsilon} P_{gg}(z)  \end{array}\right),
\end{equation}
with $\mu_h^2$ given by  Eq.~(\ref{eq:ktmaxh}), so that in the
diagonal channels the subtraction is the same as in \msbar{}, while in
the off-diagonal channels it is performed at the physical scale
$\mu_h^2$, and also, accounting for the $d$-dimensional continuation of
the  average over the polarization of the gluons.
  \end{itemize}

Positivity of the quark and gluon PDF vector $ f^{{\rm POS}}(Q^2)$,
Eq.~(\ref{eq:posmsbar2}), now follows from the same argument used to
show the positivity of the nonsinglet PDF 
Eq.~(\ref{eq:dismsbar2}). Namely, all
factors, which are convoluted in Eq.~(\ref{eq:dismsbar2}), are 
separately positive when $\epsilon \to 0^-$ and  $\mu_h^2 < \mu_D^2$ (with
$\mu_D$ defined in Eq.~(\ref{eq:mud})) and in particular, the matrix of POS-scheme coefficient
functions is now positive as shown in Sect.~\ref{sec:hadr}.

Also, as in the nonsinglet case,  the positivity argument can be formulated in terms of
a physical scheme, in which now 
to all perturbative orders the quark and gluon are defined by
\begin{equation}\label{eq:phys}
 \frac{1}{x} \bar\sigma(x,Q^2)=f^{\rm PHYS}(x,Q^2) \,,
\end{equation}
where, as in Ref.~\cite{Altarelli:1998gn}, the hadronic cross sections $\bar\sigma(x,Q^2)$ are computed
assuming that one of the two incoming protons is replaced 
by a beam of antiquarks or a beam of gluons respectively, i.e.\ 
\begin{equation}\label{eq:physxsec}
   \bar\sigma(x,Q^2)=\left(\begin{array}{c} \sigma(x,Q^2)[\bar q
       p\to \gamma^*+X]\\ \sigma(x,Q^2)[gp \to H+X]\end{array}\right).
\end{equation}
This hadronic cross section is linear in the PDFs, it coincides with it
at LO in any scheme, and, assuming that it coincides with it to all
orders, defines the PHYS scheme. Equivalently, one could choose as
$\bar \sigma$ a DIS structure function in the quark channel, and the
cross section for Higgs production in photon-gluon fusion in the gluon
channel.
The POS and PHYS schemes are then related by
\begin{equation}\label{eq:postodis}
f^{\rm PHYS}(x,Q^2)=
\left[\mathbb{I}+\frac{\alpha_s}{2\pi} {\Delta^{(1)}}^{\rm POS}
  +\frac{\alpha_s}{2\pi}  {\bar
    C^{(1)}{}}^{\rm POS}\otimes\right]f^{\rm POS}(Q^2) \,,
\end{equation}
which is perturbatively inverted as  
\begin{equation}\label{eq:distopos}
f^{\rm POS}(x,Q^2)=
\left[\mathbb{I}-\frac{\alpha_s}{2\pi} {\Delta^{(1)}}^{\rm POS}
  -\frac{\alpha_s}{2\pi}  {\bar
    C^{(1)}{}}^{\rm POS}\otimes\right]f^{\rm PHYS}(Q^2)+O(\alpha_s^2)\,.
\end{equation}

Again, this shows that positivity of the POS-scheme  coefficient function is
necessary for positivity of the POS-scheme PDFs and sufficient if
perturbativity holds. Just like in the case of
Eq.~(\ref{eq:distomsbar}), this assumption fails at the endpoints
$z\to0$ and $z\to 1$. However, as well known~\cite{Ellis:1991qj}, and
as it is easy to check from the explicit expressions of the matrix
elements of ${\bar
    C^{(1)}(z)}^{\rm POS}$, in both these limits the matrix
  is diagonal up to power-suppressed corrections. Specifically, 
 in the $z\to1$ limit the coefficient function matrix is diagonal:
\begin{equation}\label{eq:diag}
    \lim_{z\to 1}  {C^{(1)}}^{\rm POS}(z,Q^2)=\left(\begin{array}{cc} {{{C^{q}}_q}^{(1)}}^{\rm POS} &
      0 \\0 & {{{C^{g}}_g}^{(1)}}^{\rm POS} \end{array}\right)\left[1+O(1-z)\right].
\end{equation}
Indeed, diagonal coefficient functions grow as
$\left(\frac{\ln(1-z)}{(1-z)}\right)_+$ while off-diagonal ones tend to a constant
as $z\to1$. This is clearly seen in the $N$ space plots of
Figs.~\ref{fig:dy}-\ref{fig:higgs}, in which as $N\to\infty$ the diagonal coefficient
functions are seen to grow (as $\ln^2N$) while the off-diagonal ones
vanish (as $\frac{1}{N}$)~\footnote{The same power behavior also
  holds in the \msbar{} scheme, where however the off-diagonal
  coefficient functions grow as $\ln(1-z)$ as $z\to 1$, corresponding
  to a $\frac{\ln N}{N}$ behavior of its Mellin transform at large $N$.}
It follows that at large $z$ the quark and gluon channels decouple, and
the perturbativity argument is the same as in the
nonsinglet case. 

\subsubsection{Positive PDFs and their scale dependence}
\label{sec:ap}

In Section~\ref{sec:pos}
we have shown that also in the presence of quark-gluon
mixing  POS-scheme
coefficient functions are positive, and thus in the perturbative
regime PDFs are also positive. One can then ask two (closely related)
questions. First, at which scale does this conclusion apply, and is it
affected by perturbative evolution? And second, which PDF combinations
are actually positive?
Indeed, as well known, the eigenstates of QCD evolution are the two
eigenstates of a mixing matrix between the quark singlet and the
gluon,  and individual nonsinglet components; any PDF (and thus any
observable) can be decomposed into a singlet and nonsinglet component,
which evolve independently (see
e.g.\ Sect.~4.3.3 of Ref.~\cite{Ellis:1991qj}). Of course a difference
between two positive quantities is not necessarily positive, so this
raises the question of which are actually the positive combinations:
the eigenstates of evolution, or individual quark, antiquark and
gluons (or indeed something else)?

In order to answer the questions, we start from the observation that
the operators whose matrix elements separately define probability densities are
the quark operators Eq.(\ref{eq:quark}), and their antiquark and gluon
counterparts. This can be understood physically in a simple way by
considering a moment of the PDF: for example, the second moment of the
PDF for quark of flavor $i$ is just the matrix element of the energy
(Hamiltonian) operator for the corresponding quark, expressed in terms
of creation and annihilation operators for the given quark state.
Ditto for each antiquark of
flavor $j$, and for the gluon. Hence, at leading order the quantities
which are separately positive are individual quark flavors, antiquark
flavors, and the gluon.

The argument presented in Section~\ref{sec:pos} shows that this
positivity is preserved for the quark and gluon PDF, which at this
order mix to first order in $\alpha_s$. This argument does not make
any assumption about the particular value of $Q^2$, except that it
ought to be in the perturbative region where $\alpha_s(Q^2)$ is small
enough. Hence, positivity must necessarily be preserved by QCD
evolution.

Actually,  that this is the case directly follows from the construction of the
positive subtraction scheme.
Indeed, QCD evolution of the PDF is a consequence of the $Q^2$
dependence induced by the factorization into the PDF of
scale-dependent collinear logs, i.e., by the scale dependence of the
renormalization factor $Z^S_{ij}(Q^2)$ in
Eqs.~(\ref{eq:pdffac},\ref{eq:pertz}). Indeed, using in these
equations
the explicit form
of the subtraction, as given in
Eqs.~(\ref{eq:cgren},\ref{eq:cqren},\ref{eq:cgqren}) it follows that
upon a change of the scale at which the subtraction is performed, the
renormalization factor changes according to
\begin{equation}\label{eq:dglap}
  Z^S_{ij}({Q'}^2)=\left(\delta_{ij}+\frac{\alpha_s({Q'}^2)}{2\pi}P_{ij}\ln\frac{{Q'}^2}{Q^2}\right)\otimes
  Z_j(Q^2) +O(\alpha_s^2),
\end{equation}
where $P_{ij}$ is the Altarelli-Parisi splitting function. Of course,
taken in differential form for infinitesimal scale changes
Eq.~(\ref{eq:dglap}) is the standard QCD evolution equation.

The POS factorization scheme construction essentially amounts to
choosing $\delta^{S}_{ij}$ in Eq.~(\ref{eq:pertz}) in  such a way that
$Z_{ij}^{S}$ remains positive for all $Q^2$: in particular, whenever
$P_{ij}$ is negative, this will mean that as the scale is increased,
the renormalization factor $Z_{ij}^S$ decreases, while (in a positive
scheme) remaining positive. Clearly, the condition is more easily
satisfied at higher scales because of asymptotic freedom, in agreement
with the phenomenological observation~\cite{Ball:2010de,Ball:2014uwa}
that positivity constraints are more restrictive if imposed at low
scale and are preserved by evolution.

It is worth noting that a consequence of Eq.~(\ref{eq:dglap}) is that,
as well known, a scheme change will affect the NLO splitting
functions. In particular, in the POS scheme contributions proportional
to $\ln\frac{(1-z)^2}{z}$ to the off-diagonal splitting function will
now be automatically resummed to all orders when solving the NLO QCD
evolution equations. These contributions are actually power-suppressed
as $z\to1$, so this resummation is likely not to have a significant
effect: the POS scheme is thus useful as a means to obtain positive
PDFs (which is our main goal here), but not necessarily
phenomenologically better than the standard \msbar{} scheme. On the
other hand, in Ref.~\cite{Jadach:2016acv} a factorization scheme has
been advocated, called the Monte Carlo scheme, that is similar in
spirit to the POS scheme in the off-diagonal channel, but also
modifies the \msbar{} subtraction in the diagonal channel by an
analogous change of subtraction point. In this Monte Carlo scheme,
$\ln(1-z)^2$ contributions in the diagonal channels are also resummed
when solving the QCD evolution equation: hence, leading-log threshold (Sudakov)
resummation is automatically performed, without having to be added a
posteriori. It can be argued that in this Monte Carlo scheme PDFs also
resepect positivity~\cite{jadprivate}.

\subsection{Positive schemes vs.\ \msbar{}}
\label{sec:gen}

In the previous section, we have shown that coefficient functions and
PDFs in the POS
factorization scheme are indeed positive. We would like now to
investigate the relation of the POS scheme to other factorization
schemes, specifically \msbar{}, and the related issue 
of
how a positive factorization scheme should be and can be defined.

\subsubsection{General positive schemes}
\label{sec:gensch}

\begin{figure}[t]
  \begin{center}
    \includegraphics[width=0.8\linewidth]{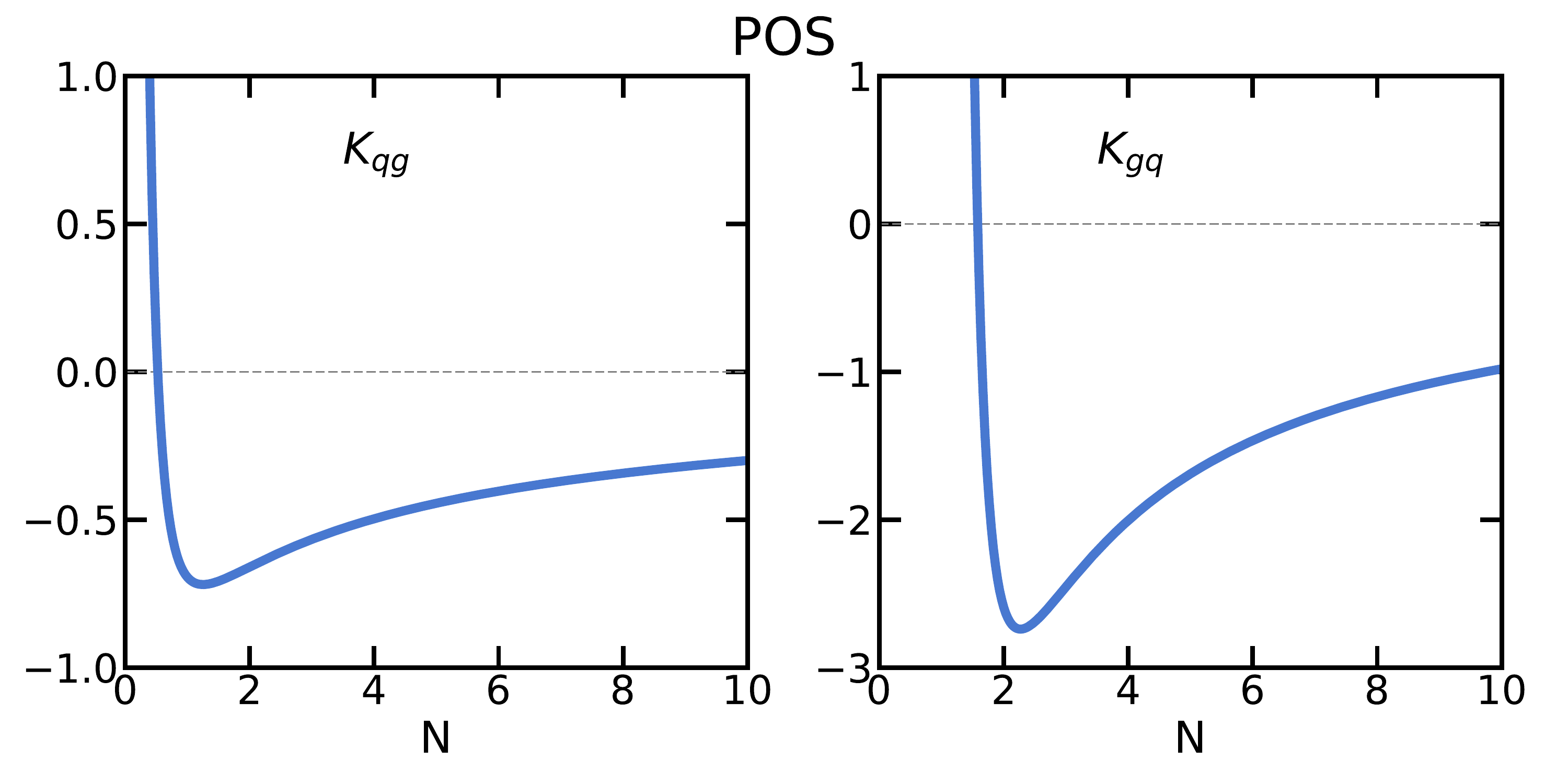}
    \caption{\small The off-diagonal  elements of the NLO scheme
      change matrix $K^{\rm POS}$, Eq.~(\ref{eq:kstruct}), in Mellin space.
    \label{fig:kmat} }
  \end{center}
\end{figure}
The scheme change from
POS to \msbar{} can be determined using Eqs.~(\ref{eq:counthqq}-\ref{eq:counthqg}) (quark channel)
and Eqs.~(\ref{eq:counthgg}-\ref{eq:counthgq}) (gluon channel).
We have
\begin{align}
  \label{eq:schch0}
\left[\mathbb{I}
+\frac{\alpha_s}{2\pi} {C^{(1)}}^{\mmsbar}\right]&=\left[\mathbb{I}
  +\frac{\alpha_s}{2\pi} {C^{(1)}}^{\rm POS}\right]\otimes\left[\mathbb{I}
  +\frac{\alpha_s}{2\pi} {C^{(1)}}^{\rm POS}\otimes\right]^{-1}\left[\mathbb{I}
  +\frac{\alpha_s}{2\pi}  \left({C^{(1)}}^{\rm POS} +K^{\rm
    POS}\right)\right]\\
&=\left[\mathbb{I}
  +\frac{\alpha_s}{2\pi} {C^{(1)}}^{\rm POS}
  \right]\left[\mathbb{I}
  + \otimes \frac{\alpha_s}{2\pi}  K^{\rm POS}\right]
  , \label{eq:schch1}
\end{align}
where in Eq.~(\ref{eq:schch1}) we have written the inverse of the
POS scheme coefficient functions in perturbative form according to
Eq.~(\ref{eq:distopos}). The matrix  $K^{\rm POS}$ has the
off-diagonal structure
\begin{equation}\label{eq:kstruct}
  K^{\rm POS}=\left[\ln\left(\frac{(1-z)^2}{z}\right) - 1\right]
  \left(\begin{array}{cc} 0 & P_{qg}(z) \\
  P_{gq}(z) & 0\end{array}\right).
\end{equation}
The off-diagonal matrix elements of the matrix are displayed in
Fig.~\ref{fig:kmat} in Mellin space. 
Writing the basic factorization formula Eq.~(\ref{eq:posmsbar2}) in
the POS and \msbar{} schemes, equating the results, and using
Eq.~(\ref{eq:schch1}) we get
\begin{align}
  \label{eq:schch}
 f^{\rm POS}(Q^2)=\left[\mathbb{I}
  +\frac{\alpha_s}{2\pi}  K^{\rm POS}\otimes\right] f^{\mmsbar}(Q^2)\,,
\end{align}
which gives the scheme change between the \msbar{} and POS PDFs. 

Inspection of Eq.~(\ref{eq:schch}) immediately shows a possible issue
with the POS scheme. Indeed, as well known, momentum conservation
implies the pair of relations between the second Mellin moments of splitting
functions $\gamma_{qq}(2)+\gamma_{gq}(2)=0$ and
 $2n_f\gamma_{qg}(2)+\gamma_{gg}(2)=0$. This relation is verified in the
\msbar{} scheme:  in order for it to remain true  in any scheme obtained
from \msbar{}, the scheme change matrix must satisfy
\begin{equation}\label{eq:momcons}
  K_{qq}+K_{gq}=2n_fK_{qg}+K_{gg}\Big|_{N=2}=0 \,,
\end{equation}
where by $K_{ij}\Big|_{N=2}$ we denote the second Mellin moment of the
scheme change matrix elements. This relation is not satisfied by the matrix defined in
Eqs.~(\ref{eq:counthqq}-\ref{eq:counthqg},\ref{eq:counthgg}-\ref{eq:counthgq}).

It might therefore be worth considering a variant of the POS scheme,
in which momentum conservation is enforced by adding to the diagonal
elements of the scheme change matrix a contribution which enforces
momentum conservation. This can be done e.g.\ by adding a soft
function, which vanishes both as $z\to1$ and $z \to0$. We choose
\begin{equation}\label{eq:fmom}
  f^{\rm MOM}(z)= 60 z^2(1-z)^2\,,
\end{equation}
which has the property that its second Mellin moment equals one:
$f^{\rm MOM}(N=2)=1$. We then define a MPOS scheme as that which is obtained
from \msbar{} through a scheme change matrix $K^{\rm MPOS}$ whose
matrix elements satisfy
\begin{align}
  \label{eq:mposqq}
  K^{\rm MPOS}_{qq}(z)&= - f^{\rm MOM}(z) K^{\rm POS}_{gq}\Big|_{N=2} \,,\\
  \label{eq:mposqg}
  K^{\rm MPOS}_{qg}(z)&= K^{\rm POS}_{qg}(z) \,,\\
  \label{eq:mposgq}
  K^{\rm MPOS}_{gq}(z)&= K^{\rm POS}_{gq}(z) \,,\\
  \label{eq:mposgg}
  K^{\rm MPOS}_{gg}(z)&= -2n_f f^{\rm MOM}(z) K^{\rm POS}_{qg}\Big|_{N=2} \,.
\end{align}
The MPOS scheme then automatically satisfies momentum
conservation. Coefficient functions in the MPOS scheme are shown in
Figs.~\ref{fig:dis}-\ref{fig:higgs}. It is clear that coefficient
functions, and thus PDFs, remain
positive in the MPOS scheme: indeed, the off-diagonal coefficient
functions are unchanged, while the diagonal NLO contributions are
modified by a small correction which is offset by the large positive
LO contribution, and in fact in the hadronic case leaves the NLO
correction positive for all $z$. Hence the MPOS and POS schemes have the same
positivity properties. We will thus not discuss the MPOS scheme
any further and restrict the discussion for simplicity to the POS
scheme.

A further observation is that the POS
scheme has been constructed in Sect.~\ref{sec:hadr} based on the kinematics of hadronic processes,
namely by performing the collinear subtraction in off-diagonal
channels at the scale  $\mu_h^2$,
Eq.~(\ref{eq:ktmaxh}). As discussed in Sect.~\ref{sec:dy}, if this
scheme is used for the computation of electroproduction processes for
which the relevant scale is $\mu_D^2$, Eq.~(\ref{eq:ktmax}), leads
to coefficient functions, and consequently PDFs, that are with stronger
reason positive. More in general, the POS scheme has been constructed
using universal properties of the collinear emission that only
depend on the LO splitting functions and the choice of scale, which is
determined by the general kinematics of hadronic processes, but
otherwise process-independent. However, the positivity argument
presented in this Section shows that this choice, whereas theoretically
appealing, is by no means necessary. In fact, any physical scheme
choice of the form of Eq.~(\ref{eq:phys}) can be used to construct a
positive factorization scheme, by just picking a scheme choice such
that the coefficient functions of the processes used to define the
PDFs remain positive, and perturbative for all $\xi$.
In any such scheme positivity of the PDFs
holds. In fact, the simplest choice would be to pick as a
positive factorization scheme the physical scheme itself, in which
PDFs are positive by construction, as they are identified with
physically observable cross sections.

\subsubsection{The \msbar{} scheme}
\label{sec:posmsbar}

Having concluded that we can take the POS scheme as representative of
a wide class of positive factorization schemes, we now discuss 
its relation to the \msbar{} scheme, and what it
tells us about positivity of \msbar{} PDFs.

Inverting the scheme change from \msbar{} to POS perturbatively (see Eq.~(\ref{eq:schch}))
we obtain
\begin{equation}
  \label{eq:schchinv}
 f^{\mmsbar}(Q^2)=\left[\mathbb{I}
  -\frac{\alpha_s}{2\pi}  K^{\rm POS}\otimes \right]  f^{\rm POS}(Q^2) \,.
\end{equation}
It is then clear that if the POS PDFs are positive, then so are the
\msbar{} ones, because the matrix $K^{\rm POS}$ vanishes on the
diagonal, and it has negative matrix elements off the diagonal, so
$-K^{\rm POS}$ in Eq.~(\ref{eq:schchinv}) is positive.
The perturbative inversion is justified due to the fact that the
non-vanishing off-diagonal matrix elements of the $K$ matrix are
actually power-suppressed (i.e.\ next-to-eikonal) in the $z\to 1$
limit.

This can be seen more formally by considering the exact Mellin-space
inverse of the scheme change matrix, Eq.~(\ref{eq:schch}):
\begin{align}\label{eq:exninv}
\left[\mathbb{I}
  +\frac{\alpha_s}{2\pi}  K^{\rm POS}(N) \right] ^{-1}=
\frac{1}{1-\left(\frac{\alpha_s}{2\pi}\right)^2 K_{qg}(N)K_{gq}(N) } \left[\mathbb{I}
  -\frac{\alpha_s}{2\pi}  K^{\rm POS}(N)\right],
\end{align}
where $K^{\rm POS}_{ij}(N)$ denote (by slight abuse of notation) the Mellin
transforms of the matrix elements $K^{\rm POS}_{ij}$ of the matrix
$K^{\rm POS}$.
It is easy to check that the factor $ K_{qg}(N)K_{gq}(N)$ is a
monotonically decreasing function of $N$ along the real $N$ axis, and
in particular it vanishes as $\frac{1}{N^2}$ as $N\to\infty$, hence
the prefactor which relates the exact and perturbative inversions,
Eqs.~(\ref{eq:schchinv}-\ref{eq:exninv}), is actually bounded in the
region $N\gtrsim 2$ in which the \msbar{} coefficient functions, and
thus the matrix elements of $K$, turn negative (see Figs.~\ref{fig:dy}-\ref{fig:higgs}). 

We conclude that the light quark and gluon \msbar{} PDFs are in fact
positive at NLO.

Heavy quarks require a separate discussion, because
for heavy quarks \msbar{} factorization can be defined in a variety of
ways (see e.g.~\cite{Forte:2010ta}).
Specifically, heavy quarks can be treated in a
massive scheme, in which   collinear singularities
associated to them are regulated by their mass, so they decouple from
perturbative evolution. In this scheme no collinear subtraction is
performed for massive quarks, so their PDF is given by the
unsubtracted 
Eq.~(\ref{eq:quark})  and thus it remains a  positive (and
scale-independent) probability distribution to all perturbative
orders. Note that nothing prevents this heavy quark PDF from having an
``intrinsic''  component, of non-perturbative origin: however, in this
factorization scheme, the heavy quark PDFs will be scale-independent,
and thus positive at all scales.

However, it is also possible to treat the heavy quark in a massless \msbar{}
scheme, in which the heavy quark is treated like other massless
quarks, namely the collinear singularity regulated by its mass is
subtracted  according to Eqs.~(\ref{eq:cgren},\ref{eq:cqren}), but with
$\mu^2$ now replaced by the heavy quark mass. Calculations performed
in this scheme, with heavy quark mass effects neglected, are accurate
for scales much larger than the quark mass.  
However,  the massless scheme is in principle formally defined for all scales,
including at the heavy quark mass. This is sometimes done by using the
massless scheme for all flavors, but discontinuously changing the
number of flavors at a matching scale chosen equal to (or of order of) the
heavy quark mass (zero-mass variable-flavor number scheme,
ZM-VFNS~\cite{Aivazis:1993pi}). Below the matching scale the ZM-VFNS
coincides with the massive scheme (with non-evolving heavy quark PDF),
and at the matching scale the heavy quark PDF
changes discontinuously:  the matching condition is the scheme transformation
from the massive to the massless \msbar{} (computed up to NNLO in
Ref.~\cite{Buza:1996wv}). This scheme transformation accounts for the
fact that in the massive scheme the heavy quark decouples from the
running, so
loop corrections with the massive
quark circulating in loops are included in the Wilson coefficient,
and not in the operator matrix element, while in the massless scheme
they are included in the operator normalization along with all other
light quarks, but neglecting the quark mass when computing them.

When $Q^2\sim m_h^2$ this neglect is not justified, and the corresponding
scheme transformation may ruin positivity of the PDF. Specifically, it
is often assumed that the massive-scheme PDF vanishes at some scale  $Q^2\sim m_h^2$,
and it indeed appears
reasonable to expect that the low-scale heavy quark scheme PDF if not vanishing, is rather smaller than
light quark PDFs (see
Refs.~\cite{Ball:2015dpa,Ball:2016neh}). However, if one determines the 
massless-scheme
heavy quark PDF by starting with a vanishing massive-scheme PDFs, and
using perturbative  matching conditions, a negative result can be found
--- and is 
indeed found using standard light quark and gluon
PDFs~\cite{Ball:2017nwa}. This is now possible because the
massless-scheme heavy quark
PDF is not defined by a matrix element of the form of
Eq.~(\ref{eq:quark}), but rather, as the transformation of such an
operator matrix element to a scheme in which the quark mass is
neglected, but in a region in which the quark mass is not negligible.
Of course, if  $Q^2\gg m_h^2$ the mass does become negligible, the
previous arguments apply, and
positivity of the heavy quark PDF is restored. Hence, positivity of
the heavy quark PDF in the messless scheme only holds at high enough
$Q^2$ that mass corrections are negligible.
 
All the discussion so far has been pursued at NLO\@. However, the main
structure of the argument remains true to all perturbative orders. In
particular,  it is true to all orders that the diagonal splitting
functions are negative at large $z$: in fact, at large $z$ to all
perturbative orders they behave as $\frac{1}{(1-z)_+}$~\cite{Albino:2000cp}. At higher
perturbative orders, coefficient functions will contain plus distributions with higher order
powers of $\ln(1-z)$, leading to  the familiar rise in the partonic
cross section which is predicted to all orders by threshold
resummation~\cite{Catani:1989ne,Sterman:1986aj}. Off-diagonal
channels, where negative contributions as $z\to1$ may and indeed are
expected to arise, 
remain power suppressed in this limit. It follows that the
off-diagonal structure Eq.~(\ref{eq:kstruct})
of the matrix relating a positive scheme to \msbar{} will hold true to
all orders. The positivity argument of Sect.~\ref{sec:posmsbar} is a
direct consequence of this structure, and it will thus also hold to
all orders.

\section{Conclusions}
\label{sec:conc}
The goal of this paper was the construction of a universal
factorization scheme in which PDFs are non-negative. In order to
attack the problem, we started from the observation that \msbar{}
partonic cross sections for typical electro- and hadro-production
processes are not positive. This then implies that positivity of the
PDFs is not guaranteed, since folding a negative partonic
cross section with a positive PDF could lead to a negative physical
cross section. We have then traced negative partonic cross sections to
the way collinear subtraction is performed in \msbar{} and specifically
we have shown that it is due to over-subtraction, related to the choice of subtraction scale, 
and also the treatment of the average over gluon polarizations in $d$
dimensions. This loss of positivity only manifests itself  in off-diagonal quark-gluon and
gluon-quark channels.

A universal subtraction prescription  which preserves positivity of the partonic
cross section can then be constructed using hadronic kinematics, and
shown to preserve positivity also in electroproduction kinematics. This
prescription does not automatically respect momentum conservation,
which however can be enforced with a soft modification of the subtraction
procedure that does not affect its positivity properties. By
performing collinear factorization in the standard approach of
Refs.~\cite{Collins:1981uw,Curci:1980uw} it is then possible to show
that positivity of the PDFs, defined  as probability
distributions, is preserved at all stages, so PDFs remain positive.

In fact, this positivity is a manifestation of the fact that PDFs can
always be defined in terms of a physical process: what PDFs do is to
allow one to express the perturbative QCD prediction for a process in
terms of that for another process. The definition of the PDFs can then
be 
process-independent (as in \msbar{}) or process-dependent (as
in so-called physical schemes~\cite{Catani:1995ze,Diemoz:1987xu}).
Its positivity will then be preserved provided only that the renormalization
conditions, which fix the value of operator matrix elements that
define the PDFs, preserves their interpretation as moments of a
probability distribution. Effectively, this corresponds to choosing
positive Wilson coefficients.

By considering a scheme in which PDFs are manifestly positive, and the
transformation from it to \msbar{}, we have finally shown that in the
\msbar{} scheme PDFs remain positive, despite the fact that off-diagonal partonic
cross sections are negative. From a physical point of view, this is a consequence
of the fact that the \msbar{} subtraction is actually strongly positive
in the diagonal channels (where by ``strongly'' we mean that
partonic  functions tend to $+\infty$  towards kinematic
boundaries). This then overwhelms the negative contribution from
off-diagonal channels, while away from kinematic boundaries
off-diagonal channels are perturbatively subleading.

Positivity of the PDFs is neither necessary nor sufficient for physical
cross sections to be positive, as they ought to: it is not necessary,
because it is possible that a negative PDF still leads to a positive
hadronic cross section once folded with a suitable coefficient
function, and it is not sufficient because in a scheme, such as
\msbar{},
in which some partonic cross sections are negative it could well be
that, while the true PDF must necessarily lead to positive measurable
cross sections,  an incorrectly determined PDF could lead to a negative
cross section despite being positive.

In other words, it is not necessarily true that 
the region in PDF space which is excluded by the requirement of
positivity of the PDF is the same as that which is excluded by
requiring positivity of the cross sections. However, from the point of
view of PDFs determination, knowing that PDFs must be
positive in a given factorization scheme does provide a useful constraint,
in that it excludes a region which does not have to be explored,
though this restriction is not necessarily the most stringent one. It
is natural to ask whether the positivity requirement could be
more restrictive in some factorization schemes than others, but it is
unclear whether and how this question could be answered. The question
of optimizing the scheme choice from the point of view of positivity
constraints, for the sake of PDFs determination, remains open for
future investigation.

\section*{Acknowledgments}
We are especially grateful to Christopher Schwan for a careful
critical reading of the manuscript,  to Richard Ball,
Zahari Kassabov, Luca Rottoli and  Maria Ubiali  for numerous comments
and criticisms on a  preliminary version of the paper, to Tommaso
Giani and Rabah Abdul Kaleh for  questions and critical input, to
Rosalyn Pearson for a thorough revision of the draft, and to Stanislaw
Jadach for discussions and correspondence.
This work is  supported by the European Research Council under
the European Union's Horizon 2020 research and innovation Programme
(grant agreement n.740006).

\bibliographystyle{UTPstyle}
\bibliography{positivity}

\end{document}